\documentclass{nature}

\usepackage{graphicx,epsfig}% Include figure files
\usepackage{amsmath,amssymb,mathtools}%,amsfonts,multirow,rotate,color}amssymb
\usepackage[rgb]{xcolor}
\usepackage{hyperref}
\usepackage{comment}

\def\be{\begin{equation}}
\def\ee{\end{equation}}

\def\bc{\begin{center}}
\def\ec{\end{center}}
\def\bea{\begin{eqnarray}}
\def\eea{\end{eqnarray}}
\newcommand{\avg}[1]{\langle{#1}\rangle}
\newcommand{\Avg}[1]{\left\langle{#1}\right\rangle}

\def\ie{\textit{i.e.}}

\makeatletter
\let\saved@includegraphics\includegraphics
\AtBeginDocument{\let\includegraphics\saved@includegraphics}
\renewenvironment*{figure}{\@float{figure}}{\end@float}

\makeatother

\title{The dynamic nature of percolation on networks with \\ triadic interactions}

\author{Hanlin Sun$^{1}$, Filippo Radicchi$^{2}$, J\"urgen Kurths$^{3,4}$ \& Ginestra Bianconi$^{1,5}$}

\begin{document}

\maketitle

\begin{affiliations}
 \item School of Mathematical Sciences, Queen Mary University of London, London E1 4NS, United Kingdom
 \item Center for Complex Networks and Systems Research, Luddy School
  of Informatics, Computing, and Engineering, Indiana University, Bloomington, %IN
  47408, USA
 \item Potsdam Institute for Climate Impact Research, Potsdam, Germany
\item Department of Physics, Humboldt University of Berlin, Berlin, Germany
 \item The Alan Turing Institute, The British Library, London NW1 2DB, United Kingdom
\end{affiliations}

\newpage

\section*{Abstract}

\begin{abstract}
Percolation establishes the connectivity of complex networks and is one of the most fundamental critical phenomena for the study of complex systems.  
   On simple networks, percolation
  displays a second-order phase
  transition; on multiplex networks, the percolation
  transition can become discontinuous. However, little is known about percolation in networks with higher-order interactions. Here, we show that percolation can be turned into a fully fledged dynamical process when higher-order interactions are taken into account. By introducing signed triadic interactions, in which a node can regulate the interactions between  two other nodes, we define triadic percolation.  We uncover  that in this paradigmatic model the connectivity of the network changes in time and that the order parameter undergoes a period doubling and a route to chaos. We provide a general  theory for triadic percolation which accurately predicts the full phase diagram on random graphs as confirmed by extensive numerical simulations. We find that triadic percolation on real network topologies reveals a similar phenomenology. These results radically change our understanding of percolation and may be used to study complex systems in which the functional connectivity is changing in time dynamically and in a non-trivial way, such as in neural and climate networks.
 \end{abstract}

\newpage

    \section{Introduction}  
    Percolation\cite{dorogovtsev2008critical,stauffer1992introduction,li2021percolation,araujo2014recent} is one of the most fundamental critical phenomena defined on networks.
    As such, it
    has  attracted large interest in the literature\cite{buldyrev2010catastrophic,goh2014network,kahng009percolation,boettcher2012ordinary,d2019explosive,achlioptas2009explosive,riordan2011explosive,da2010explosive,nagler2020universal,cho2013avoiding}. Indeed by predicting the size of the giant component (GC) of a network  when links are randomly damaged, percolation can be used  for the establishment of the minimal requirements that a structural network should satisfy in order to support any type of interactive process. 
Despite the great success of percolation, ordinary percolation is unsuitable to describe real-world situations that occur in neuronal and climate networks {  when} the connectivity of these networks changes in time. 

Typically, the dynamics associated to percolation is the one of
{   a cascading process where an initial failure propagates within a network possibly affecting its macroscopic connectedness.}
In the last decade, large scientific activity has been addressed to generalized percolation problems that capture cascades of failure events  \cite{buldyrev2010catastrophic,baxter2012avalanche,radicchi2015percolation,radicchi2017redundant,reis2014avoiding,kryven2019bond,gao2012networks} 
 on multilayer networks~\cite{bianconi2018multilayer,boccaletti2014structure,kivela2014multilayer}  where the damage propagates back and forth among the layers reaching a steady state at the end of the cascading process. In duplex networks, period-two oscillations can be observed  in presence of  competitive or antagonistic interactions~\cite{zhao2013antagonistic,danziger2019dynamic,shekhtman2016recent,watanabe2016resilience,kotnis2015percolation} among the different layers of the multiplex networks.
However, this phenomenon seems to be restricted to duplex networks.
Finally in damage and recovery models on multilayer networks \cite{majdandzic2016multiple,danziger2022recovery,shekhtman2016recent} aimed at getting insight for the robustness of  complex  critical infrastructures and financial systems, also more than two coexisting stable configurations of percolation have been observed. 

An important question that arises from these works is whether percolation can capture more general time-dependent variations in the connectivity of a network.
Here,  we give a positive answer to this question and we show  that higher-order interactions, and specifically triadic interactions, can turn percolation into a fully fledged dynamical process {   in which
  the order parameter undergoes period doubling and a route to chaos}.

Higher-order networks are ubiquitous in nature~\cite{battiston2020networks,bianconi2021higher,perc2022dynamics,lambiotte2018simplicial, bick2021higher,torres2021and}. Paradigmatic examples are the networks that describe brain activity, chemical reactions networks,  and climate~\cite{giusti2016two,faskowitz2022edges,jost2019hypergraph,boers2019complex,su2022climatic}.
Higher-order interactions may profoundly change the physical properties of a dynamical process compared to those displayed by the same process occurring on a classic network of pairwise interactions. Examples include 
synchronization~\cite{millan2020explosive,skardal2019abrupt,zhang2021unified,mulas2020coupled}, random walk dynamics~\cite{carletti2020random}, 
contagion dynamics~\cite{st2021universal,de2020social,iacopini2019simplicial,ferraz2021phase,sun2021higher,taylor2015topological}
and game theory~\cite{alvarez2021evolutionary}. However little is know so far about percolation in presence of higher-order interactions\cite{bianconi2018topological,sun2021higher,lee2021homological,bianconi2019percolation,bobrowski2020homological,bao2022impact}.

In this paper, we focus on a paradigmatic type of higher-order interactions named triadic interactions which occur when a  node regulates
the interaction between two other nodes.
Regulation can be either positive, in the sense that the node facilitates the interaction, or negative, meaning that the regulator inhibits the interaction.
Triadic interactions occur in ecosystems, where the competition between two species can be affected by the presence of a third species~\cite{kishony2016high,grilli2017higher,stouffer2019mechanistic}.
In neuronal networks, the interactions between  neurons/glia is known to be triadic with
% the
glias modulating the synaptic interaction between neurons\cite{cho2016optogenetic}.
In climate networks of extreme rainfall events, triadic interactions can be used to explain the situations in which the network links are modulated by large-scale patterns, such as
Rossby waves, which have a regulatory activity on climate inducing long-range synchronization of rainfall between Europe, Central Asia and even East Asia \cite{boers2019complex}.
Finally in chemical reaction networks, generalized triadic interactions could model the action of  enzymes as biological catalysts for biochemical 
reactions.   
While triadic interactions have received  large attention in ecology and neuroscience,  theoretical analyses of triadic interactions have investigated exclusively   
small-scale ecological systems~\cite{kishony2016high,grilli2017higher,stouffer2019mechanistic}.

Here, we change perspective and
%we
study the role of  triadic interactions in shaping macroscopic network properties. Specifically, we investigate how triadic interactions can change the critical and the dynamical properties of percolation.
We combine percolation theory\cite{dorogovtsev2008critical,li2021percolation} with the theory of dynamical systems\cite{arenas2008synchronization,kurths2009complex,strogatz2018nonlinear,porter2016dynamical} 
to define triadic percolation, i.e., percolation in presence of  signed triadic interactions.
We show that in triadic percolation the GC of the network displays a highly non-trivial dynamics
characterized by
period doubling and a route to chaos. 
We use a general theory to demonstrate that the phase diagram of triadic percolation has fundamental differences with the phase diagram of ordinary percolation.
While ordinary percolation displays a second-order phase transition, the phase diagram of triadic percolation is much richer and can be interpreted as an orbit diagram for the order parameter.
Our theory is validated with extensive simulations on synthetic and real-world networks. These results reveal that in triadic percolation the GC of the network becomes a dynamical entity whose dynamics changes radically our understanding of percolation.

\section{Results}

      \begin{figure}[!htb]
      \begin{center}
  \includegraphics[width=1\textwidth]{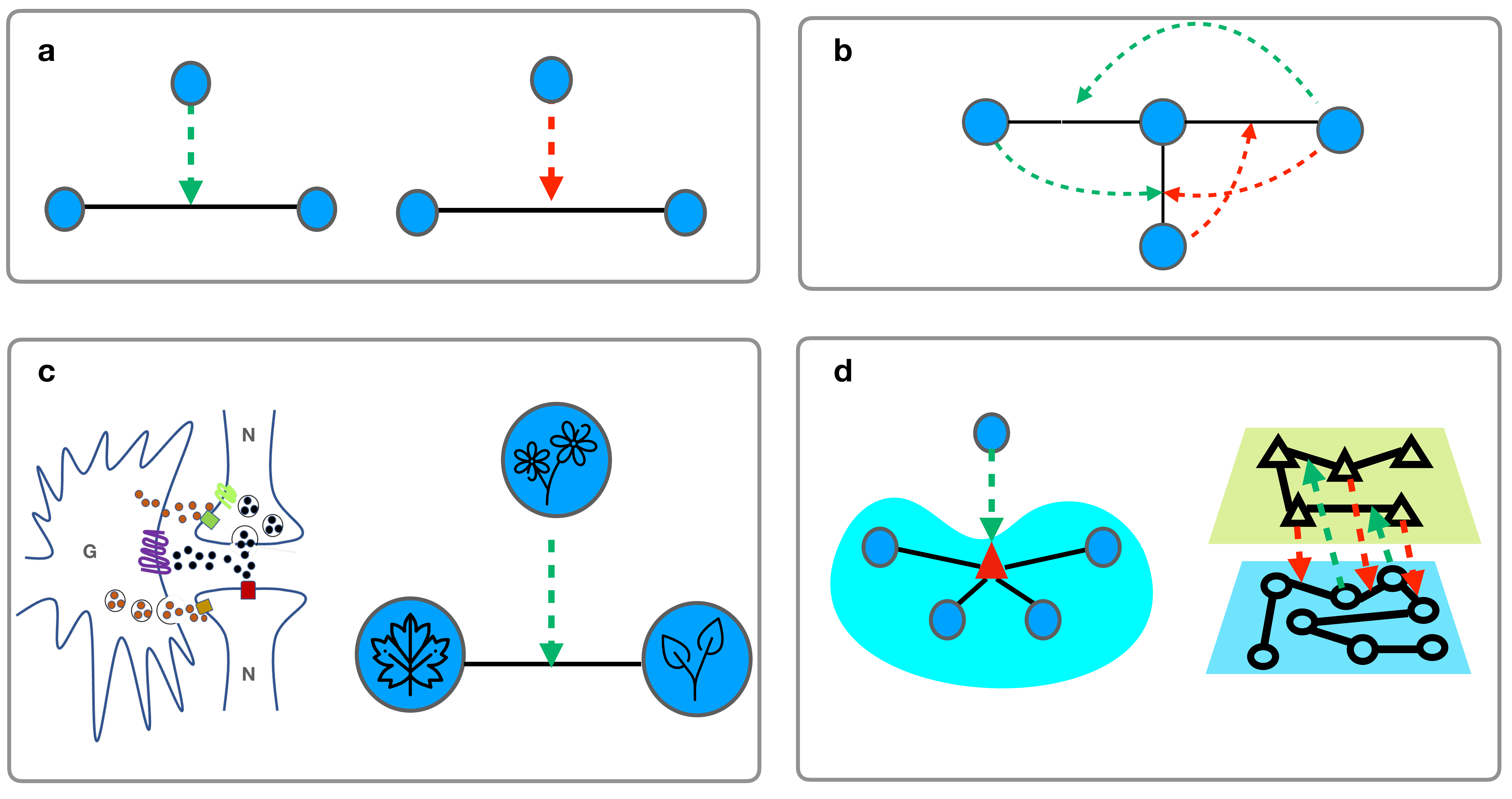}
  \end{center}
  \caption{{\bf Triadic interactions.}  Triadic interactions occur when a node regulates the interactions between  two other nodes. Triadic interactions can be signed with one node either favoring (green dashed link) or inhibiting (red dashed link) the interactions between the other to nodes (panel a). The simplest network including triadic interactions (panel b) is formed by a structural network between nodes and (solid line) structural links and a regulatory network including the regulatory interactions (dashed lines) between nodes and structural links. Examples of triadic interactions (panel c) include glias/neurons interactions and interactions between species in ecosystems. Triadic interactions can be extended to hypergraphs and multiplex networks (panel d). In hypergraphs the triadic interactions can regulate the presence or the activity of an hyper-edge, in multiplex networks triadic interactions can be used to establish inter-layer interactions between nodes in one layers and links in the other layer. 
The plant icons are made by freepik from www.flaticon.com.}
  \label{figure1}
 \end{figure}

\begin{figure}
\centering
  \includegraphics[width=0.4\textwidth]{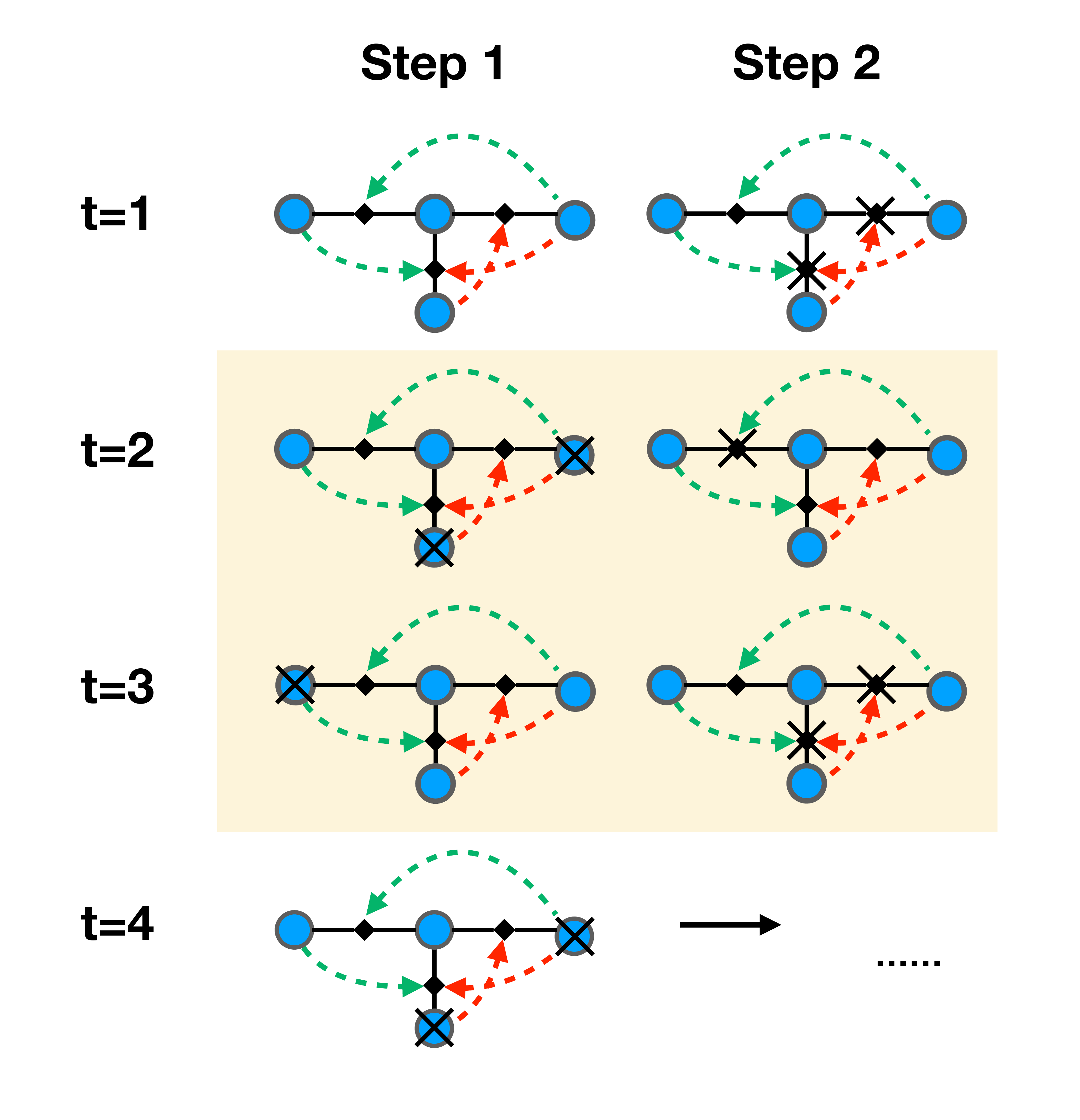}
  \caption{{\bf Sketch of triadic percolation.} Solid lines represent structural links, dashed curves denote regulatory interactions (green stands for positive regulation, red for negative). Blue filled circles indicate structural nodes, black diamonds indicate triadic interactions. 
  For simplicity, we consider the deterministic bond-percolation model for $p=p_0=1$. 
  At each stage $t$ of the dynamics, bond percolation is applied to the network, and then the effect of the regulatory activity is established.
  The illustration shows how the dynamics sets into a periodic pattern with the giant component of the network ``blinking" in time. 
  The periodic pattern is highlighted in yellow. 
  At time $t=1$, all links are active and all nodes are part of the giant component (GC). Their regulatory activity causes some links to become inactive (crossed links in the figure). As a consequence, at time $t=2$, some nodes are no longer part of the GC and become inactive (crossed nodes in the figure). However, this change leads to changes in the activity of some links, which in turn affect the activity of the nodes at time $t=3,4$, etc. The final configuration reached at time $t=3$ is identical to one observed at the end of stage $t=1$. Due to the determinism of the model, the pattern repeats with period $T=2$. The relative size $R$ of the GC oscillation switches between $2/4$ and $3/4$.   {For an example of  more complex dynamical behaviour see Supplementary movie.}
  }
  \label{figure2}
  \end{figure}
  
  {\it Triadic interactions -}
Triadic interactions (see Fig.~$\ref{figure1}$) are higher-order interactions between nodes and links. They occur when a node regulates the interaction between  two other nodes.
The regulation can be either positive, in the sense that the node facilitates the interaction, or negative, meaning that the regulator inhibits the interaction. 
For instance, the presence of a third species can enhance or
%can
inhibit the interaction between two species;
% and
also, 
the presence of a glia can favor or
% can
inhibit  the synaptic interactions between two neurons.
Triadic interactions can be
% introduced in addition to
added to
a simple structural network. However, triadic interactions can also be introduced on top of an hypergraph, when one node regulates the strength of an hyperedge, or on top of multilayer networks, where triadic interactions represent inter-layer interactions between the nodes of one layer and the links of other layer. For instance, an enzyme is a node that can regulate an hyperedge (i.e., a reaction between chemicals); neural networks and networks of glias form instead two layers of a multiplex network interacting via triadic interactions.

Let us now  formulate the simplest example of higher-order networks with triadic interactions.
This higher-order network can be modelled as the  composition of two networks: the structural network and the regulatory network which encodes triadic interactions. The  structural network $\mathcal{A}=(V,E)$ is formed by the set of  nodes $V$ connected by the structural links in the set $E$. The regulatory network $\mathcal{B}=(V,E,W)$ is a bipartite, signed  network  between  the set of  nodes $V$ of the structural network and the set of structural links $E$, with nodes in $V$ regulating links in $E$ on the basis of the regulatory interactions, either positive or negative, specified in the set $W$.  Given a regulated link, a node at the end of the regulatory interaction is called positive regulator if the regulatory interaction is positive and negative regulator if the regulatory interaction is negative.  Note that the sign is an attribute of the regulatory interaction and not of the node that acts as regulator.

In the following we will focus on percolation on this model of network with triadic interactions, however our results can be easily extended to hypergraphs and multiplex networks with triadic interactions as well.

{\it Triadic percolation -} We define triadic percolation as the model in which the activity of the structural links is regulated  by the triadic interactions and the activity of their regulator nodes. Conversely, the activity of the nodes is dictated by the connectivity of the network resulting after considering only the active links. In particular, we assume that the activity of nodes and links is changing in time leading to the triadic percolation process defined as follows. 
At time $t=0$, every structural link is active with probability $p_0$. We then iterate the following algorithm for each time step $t \geq 1$:
 \begin{itemize}
 \item[Step 1] Given the configuration of activity of the structural links at time $t-1$, we define each node active if the node belongs to the GC of the structural network in which we consider only active links. The node is considered inactive otherwise.
 
 \item[Step 2] Given the set of all active nodes obtained in step 1, we  deactivate all the links that are connected at least to one active negative regulator node and/or that are not connected to any active positive regulator node. All the other links are deactivated with probability $q=1-p$.
 \end{itemize}

 Note that for $p=p_0=1$ the model is deterministic. 
 However, for $p<1$ (and $p_0<1$) the model is stochastic
 , i.e., {   the activity of the nodes does not uniquely define the activity of the links}. 
 
 In the proposed triadic percolation, links can be dynamically {   turned} on and off by the regulatory interactions. The model only makes minimal and justifiable assumptions while remaining general. The assumption that only nodes within the GC of the network are considered functioning/active is well accepted in the literature concerning network robustness\cite{buldyrev2010catastrophic,dorogovtsev2008critical}. 
 Also, the regulatory  rule chosen for deactivating the links is the minimal rule for treating both  positive/negative regulations in a symmetric way: given suitable conditions the activation of a single positive regulator or the deactivation of a single negative regulator can turn the activity of a link on. Finally, the introduction of annealed stochastic effects, {  present for $p<1$ (and $p_0<1$),
 represents a simple way to account for the unavoidable randomness that can affect the activation/deactivation of the  structural links in real scenarios.}

Triadic percolation can lead to a highly  non-trivial dynamics of the  network connectivity. For instance  Fig.~$\ref{figure2}$ illustrates the phenomenon of 
network ``blinking" with nodes of the network turning on and off periodically to form GCs of different size.
As we will see, this dynamics
emerges
at the bifurcation transition indicating the onset of the period-two oscillations of the order parameter, but oscillations of longer period and also chaos is  observed depending on the model's parameters.
\begin{figure}
  \includegraphics[width=0.9\textwidth]{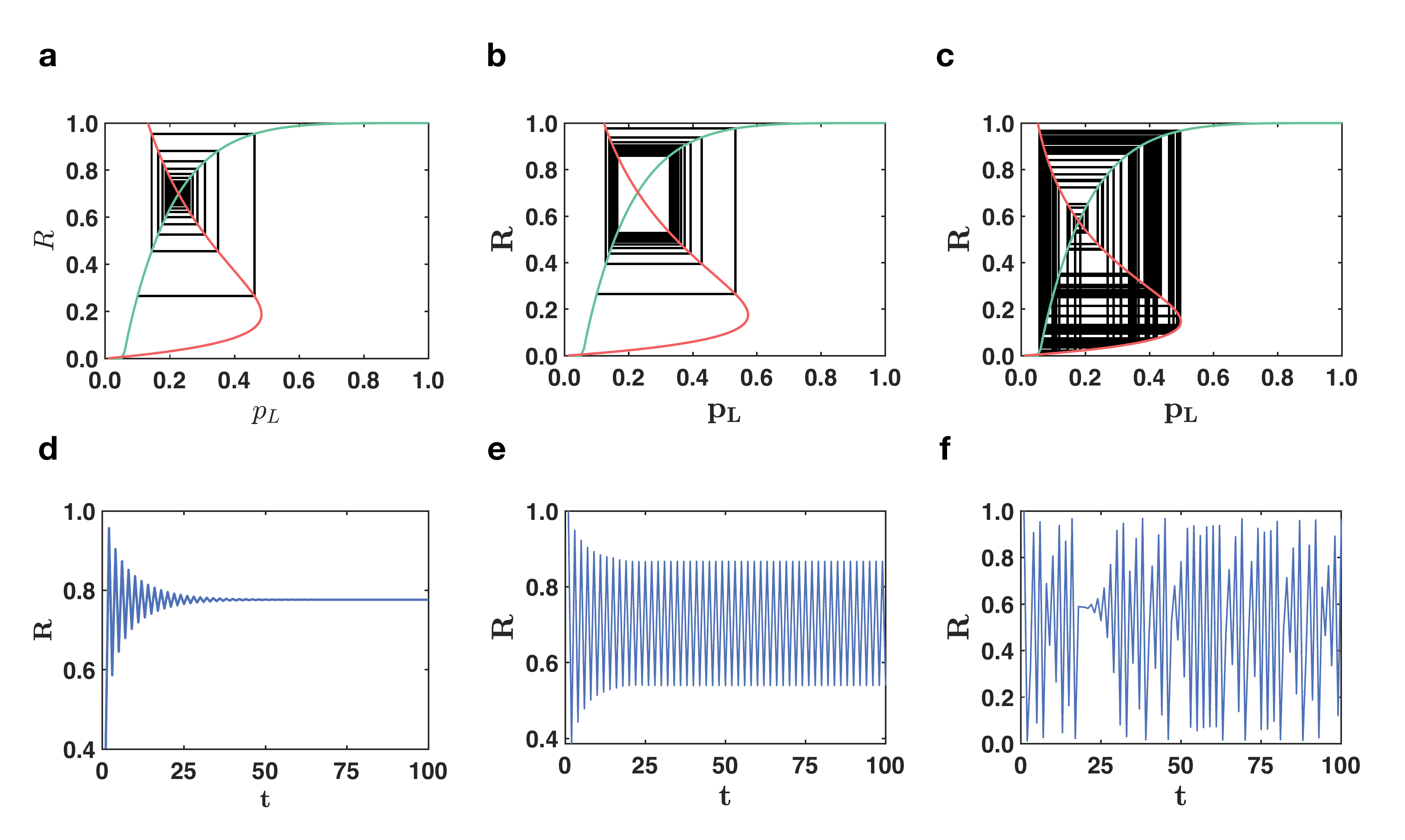}
  \caption{{\bf Time dependence of the order parameter of triadic percolation.} In triadic percolation, the order parameter $R$ can have non-trivial dynamics. Here we demonstrate with theory and simulations the  non-trivial dynamics of $R$ for parameters values in which the dynamics reaches a steady state (panels a, d), period-two oscillations (panels b, e) and a chaotic dynamics (panels c, f).
  This behaviour is predicted by the theory which can be schematically represented by  cobweb plots (panels a-c) corresponding to the map Eq. (\ref{eq:cobweb}) with the function $f$ indicated in green and the function $g_p$ in red. Results of Monte Carlo simulations for  $R$ as a function of time $t$ (panels d-f) are in excellent agreement (MC) with the theory.
   The structural network has a power-law degree distribution $\pi(k) \sim k^{-\gamma}$, with minimum degree $m=4$, maximum degree $K=100$, and degree exponent $\gamma=2.5$. The degrees  $\hat{\kappa}^{+}$ and $\hat{\kappa}^{-}$ of the regulatory network obey Poisson distributions with average ${c}^{+}$ and $c^-$   The links are activated with probability $p=0.8$. The parameters $c^+,c-$ are $c^+=10,$ $c^-=1.8$ (panel a, d), $c^+=10, c^-=2.1$ (panel b, e). The MC simulations are performed an  networks of  $N=10^4$ nodes. 
  }
  \label{figure3}
 \end{figure}
  
{\it Theory of triadic percolation -} Here we establish the theory for triadic percolation that is able to predict the phase diagram of the model on random networks with triadic interactions.

 We assume that the structural network $\mathcal{A}$ is given and contains $N$ nodes and $\avg{k} N/2$ structural links, with $\avg{k}$ indicating the average degree of the network. 
 We consider structural networks given by individual instances of the  configuration model. To this end, we first generate degree sequences by selecting random variables from
 {the degree distribution $\pi(k)$. We denote with $k_i$ the structural degree of node $i$.}

To generate the regulatory network $\mathcal{B}$, we assume that
every node $i$ has associated 
two degree values, namely
the number of positive regulatory interactions $\kappa_i^{+}$, and the number of negative regulatory interactions $\kappa_i^{-}$. For simplicity we consider the case in which  both $\kappa_i^{+}$ and $\kappa_i^{-}$ are chosen independently of the structural degree $k_i$
{   (see the SI for the extension to the correlated case).}
Each structural link $\ell$ is assigned the degrees $\hat{\kappa}_\ell^{+}$ and $\hat{\kappa}_\ell^{-}$ indicating the number of positive regulators and the number of negative regulators, respectively.  In particular, nodes' degrees are extracted at random from the distribution 
$P(\kappa^{+},\kappa^{-})$,
and links' degrees are randomly extracted from the distribution {   $\hat{P}(\hat{\kappa}^{+},\hat{\kappa}^{-})$ here taken to be uncorrelated so that  $\hat{P}(\hat{\kappa}^{+},\hat{\kappa}^{-})=\hat{P}_+(\hat{\kappa}^{+})\hat{P}_-(\hat{\kappa}^{-})$.} Once degrees have been assigned to nodes and links, we establish the existence of a positive ($+$) or negative ($-$) regulatory interaction between the structural link $\ell$ and the node $i$ with probability
\bea
p_{\ell,i}^{\pm}=\frac{\kappa_i^{\pm} \hat{\kappa}_{\ell}^{\pm}}{\avg{\kappa^{\pm}}N} \; ,
\eea
where $\avg{\kappa^{\pm}}$ denotes the average of $\kappa$ over all the  nodes of the network.  In the creation of regulatory interactions, we allow any pair  $(\ell,i)$ to be connected either by a positive of by a negative regulatory interaction but not by both. Note that as long as the network $\mathcal{B}$ is large and  sparse the latter condition is not inducing significant correlations.

Let us now combine the theory of percolation with the theory of dynamical systems to derive the phase diagram of  the considered  uncorrelated scenario.
Let us define $S^{(t)}$ as the probability that a node at the endpoint of a  random structural link of the network ${\mathcal {A}}$ is in the GC at time $t$. Moreover, let us indicate by  
$R^{(t)}$ the fraction of nodes in the GC at time $t$ (or equivalently the probability that a node at the end of a regulatory link is active).
Finally,
$p_L^{(t-1)}$ is the probability that a random   structural link is   active at time $t$. By putting $p_L^{(0)}=p_0$ indicating the probability that structural links are   active at time $t=0$, we have that for $t>0$, as long as the network is locally tree like, $S^{(t)}$, $R^{(t)}$ and $p_L^{(t)}$ are updated as
\bea
S^{(t)}&=&1-G_1\left(1-S^{(t)} p_L^{(t-1)}\right),
%\quad
\nonumber \\
{R}^{(t)}&=&1-G_0\left(1-S^{(t)} p_L^{(t-1)}\right),\nonumber \\
p_L^{(t)}&=& p  G_0^{-}(1-{R}^{(t)}) \left[1-G_0^{+}\left(1-{R}^{(t)}\right)\right],
\label{reg_percolation}
\eea
where the first two equations implement Step 1, i.e.,
{    a bond-percolation model\cite{dorogovtsev2008critical}
% when
where
links are retained with probability $p_L^{(t-1)}$,} and the third equation implements Step 2, i.e., the regulation of the links.
Here the generating functions $G_0(x),G_1(x)$ and ${G_0}^{\pm}(x)$ are given by
\bea
&G_0(x)=\sum_{k}\pi(k)x^k, \quad
%\nonumber \\
G_1(x)=\sum_{k} \pi(k)\frac{k}{\avg{k}}x^{k-1},\nonumber \\
&{G_0}^{\pm}(x)=\sum_{{\kappa}_{\pm}}\hat{P}_{\pm}({\hat{\kappa}}^{\pm})x^{{\hat{\kappa}}^{\pm}}.
\eea
Eq.~(\ref{reg_percolation}) for the percolation model regulated by triadic interactions can be formally written as the map \cite{strogatz2018nonlinear}:
\bea
R^{(t)}=f\left(p_L^{(t-1)}\right), \quad
p_L^{(t)}=g_p\left(R^{(t)}\right),
\label{eq:cobweb}
\eea
which can be further reduced to a unidimensional map $R^{(t)}=h(R^{(t-1)})$.
The previous set of  equations lead to the theoretical prediction for triadic percolation defined on structural networks generated according to the configuration model. This solution are of mean-field nature: while triadic percolation dynamics has many interacting degrees of freedom given by the activity of each node and each link, and is characterized by a stochastic dynamics for $p<1$, Eqs.~(\ref{reg_percolation}) [or equivalently the map Eqs. (\ref{eq:cobweb})] involve only three/two  variables and are deterministic.
As we will see, despite this approximations, the proposed theoretical approach provides a very accurate prediction of the behaviour of triadic percolation.

   \begin{figure}
   \begin{center}
  \includegraphics[width=0.7\textwidth]{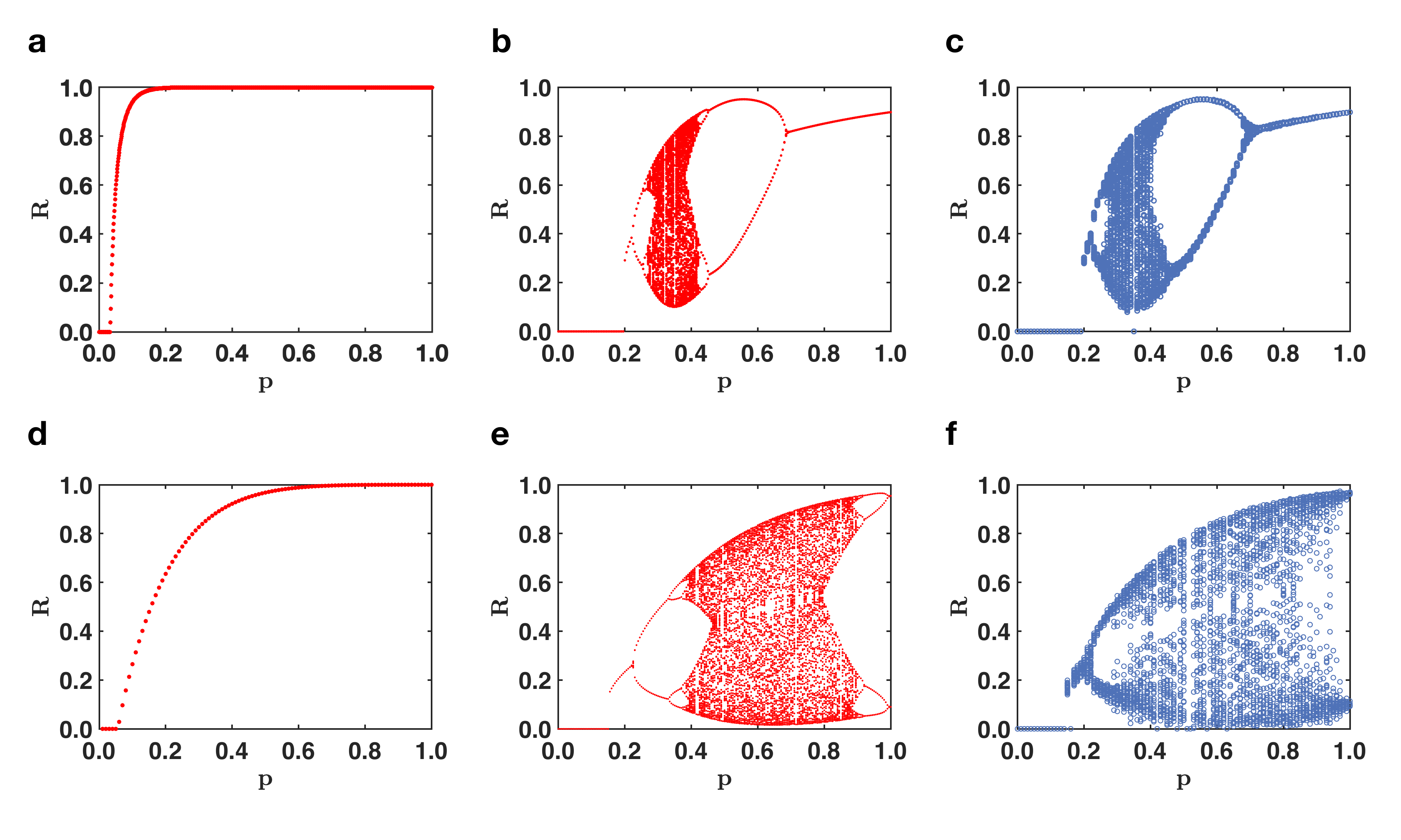}
  \end{center}
  \caption{{\bf Phase diagram of triadic percolation on Poisson and scale-free structural networks.}
  The phase diagram of triadic percolation (panels b, c, e, f) is radically different from the phase diagram of ordinary percolation (panels a, d) for both Poisson (panels a-c) and scale-free structural networks (panels d-f). Ordinary percolation reveals a second order phase transition (theoretical prediction, panels a, d) while the phase diagram of triadic percolation  reveals that the order parameter $R$  displays period doubling and a  route to chaos (panels b, c, e, f). The theoretical predictions of the phase diagram  obtained from Eq.~(\ref{reg_percolation}) are in very good agreements with the phase diagram obtained from extensive Monte Carlo (MC) simulations (panels e, f).
 In panels (a-c) the  structural network is Poisson with average degree $c=30$; the  regulatory network is also Poisson with averages $c^+=1.8$ and $c^-=2.5$.
 In panels (d-e) the scale-free structural network has degree exponent $\gamma=2.5$, minimum degree $m=4$ and maximum degree $K=100$; the regulatory network is Poisson with $c^+=10$ and $c^-=2.8$.
The  MC simulations are obtained from  networks of size  $N=2 \times 10^5$ (panel e) and $N=10^4$ (panel f). Here  points represent all $R$ values observed in the time range $150 \leq t \leq 200$.  }
  \label{figure4}
 \end{figure}
 In presence of negative interactions, triadic percolation {  displays a time-dependent } order parameter, given by the active fraction of nodes $R^{(t)}$. {   The order parameter $R^{(t)}$  undergoes a period doubling and a route to chaos in the universality class of the logistic map for
structural networks with arbitrary degree distribution $\pi(k)$ and  
 regulatory connectivity  generated by
Poisson distributions $P(\hat{\kappa}^{\pm})$
(see SI and Supplementary Figs. 1-5 for details)}. Triadic percolation has a very rich dynamical nature and displays the emergence of  both ``blinking" oscillations and chaotic patterns of the giant component (see Fig.~$\ref{figure3}$). ``Blinking" refers to the intermittent switching on and off of two or more sets of nodes which leads to periodic oscillations of the order parameter.  Chaos implies that at each time a different set and number of nodes
is activated. The map defined by  Eq.~(\ref{eq:cobweb})  allows us to generate the cobweb of the dynamical process. Theoretical predictions display excellent agreement with  extensive simulations of the model (see Fig.~$\ref{figure3}$).
  \begin{figure}
  \includegraphics[width=\textwidth]{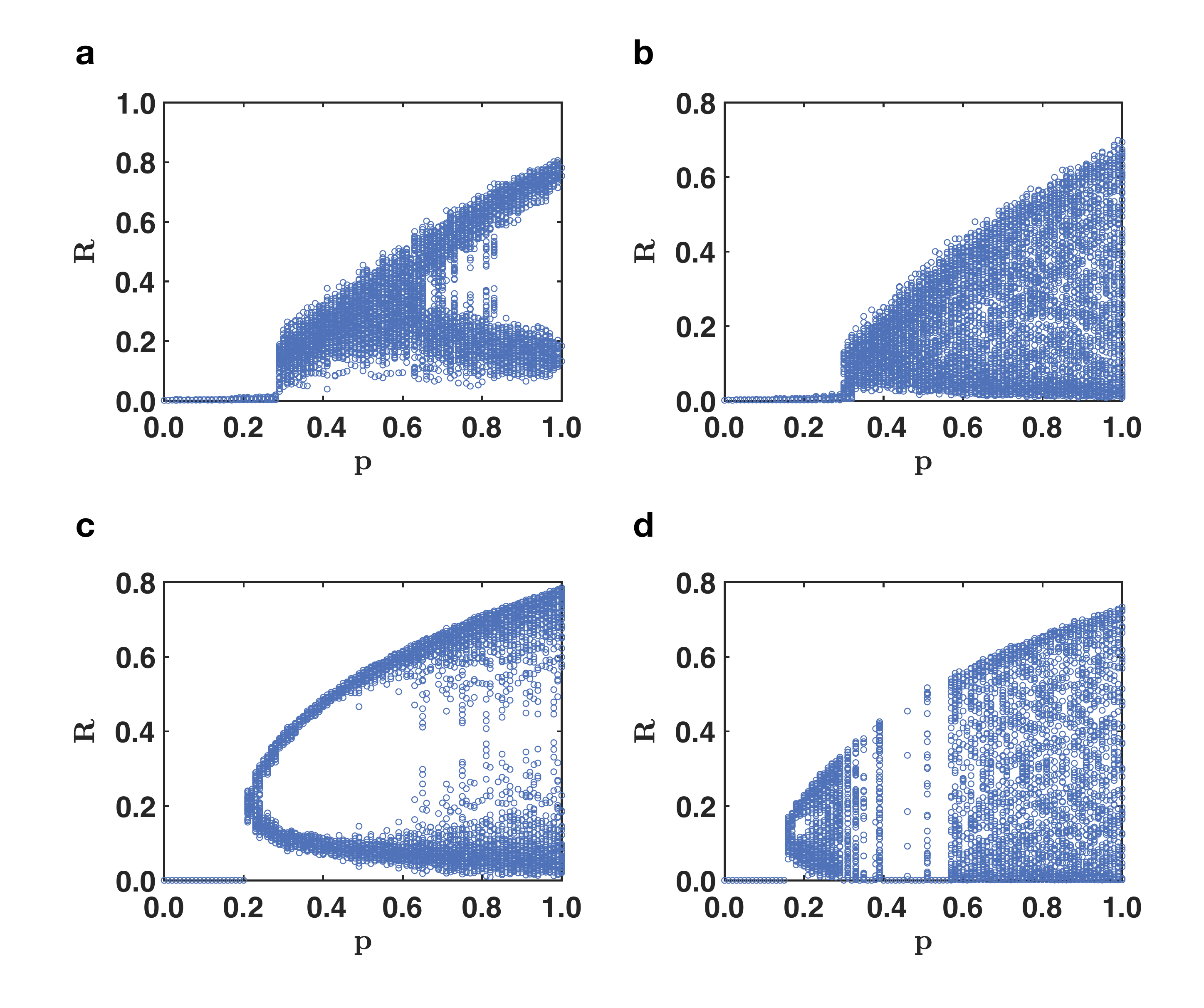}
  \caption{{\bf Phase diagram of triadic percolation for real-world structural network topologies.}
  The phase diagram  of triadic interaction displaying the fraction of nodes $R$ in the GC as a function of $p$ is shown for real-world  structural networks obtained from the repository ~\cite{Repository_networks}: the mouse brain network (panel a, b) the Human bio grid network (panel c, d).
The phase diagrams are obtained by MC simulations  with Poisson regulatory networks  with parameters $c^+=20,c^-=2$ (panel a), $c^+=20,c^-=4$ (panel b); $c^+=20,c^-=4$ (panel c).  $c^+=20,c^-=6$ (panel d). All orbit diagrams are obtained with an initial condition $p_L^{(0)}=0.1$.
  }
  \label{figure5}
 \end{figure}
 The combination of negative and positive regulatory interactions present in triadic percolation leads to a much richer phase diagram  than the one of ordinary percolation in absence of triadic interactions (see Fig.~$\ref{figure4}$).  The phase diagram of triadic percolation is found by monitoring the relative size $R$ of the GC as a function of the parameter $p$ indicating the  probability  that a link is active  when all the  regulatory conditions allowing the link to be active are satisfied. Clearly from Fig.~$\ref{figure4}$, we see that while in absence of triadic interactions the transition is second order; when signed positive and negative regulatory interactions are taken into account, the phase diagram of percolation becomes an orbit diagram.
In particular, Eq.~(\ref{reg_percolation}) predicts that the order parameter undergoes a period doubling and a route to chaos irrespective of the degree distribution of the structural network. Theoretical predictions are well matched by results of numerical simulations (see Fig.~$\ref{figure4}$). Our theory allows  to well approximate the dynamical behaviour of triadic  percolation  for random Poisson and scale-free structural networks (see Supplementary Information (SI) and Supplementary Figs. 6-11 for a discussion about the effect of the structural degree distribution on the phase diagram of triadic percolation).  

Results of numerical simulations denote a rich dynamical behaviour of the model also if structural networks are taken from the real world. In particular,  we consider real-world structural networks constructed from empirical data collected in the repository of  Ref.~\cite{Repository_networks}, and  we combine these real structural networks with synthetic regulatory networks capturing the triadic interactions.  In Fig.~$\ref{figure5}$, we show that also for these topologies the phase diagram  reveals non-trivial dynamics with some regimes of (noisy) oscillations and some regimes of chaotic dynamics of the order parameter (for more information about these datasets see Supplementary Table 1 and Supplementary Fig. 12).

\begin{figure}[!htb]
  \includegraphics[width=\textwidth]{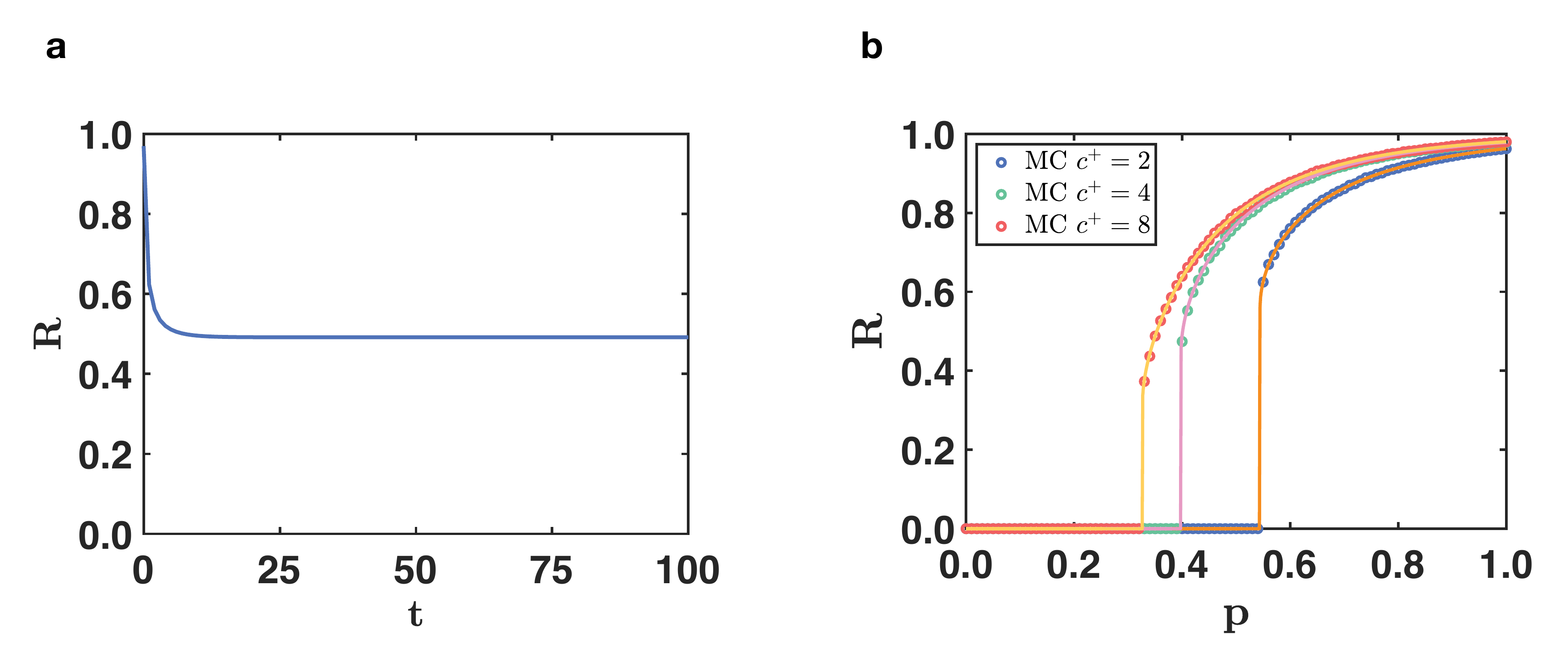}
  \caption{{\bf Triadic percolation in absence of negative triadic interactions.} In absence of negative triadic interactions the order parameter $R$ of triadic percolations always reaches a stationary state for sufficiently long times (panel a). Moreover the phase diagram, indicating the stationary solution of $R$ as a function of $p$ displays a discontinuous hybrid transition (panel b).  In panel b the results obtained from  MC simulations (symbols) over networks of  $N=10^4$ nodes are compared to theoretical  expectations (solid curves).
   In both plots the Poisson structural network has average degree $c = 4$, the Poisson regulatory network including exclusively positive regulations has average degree ${c}^{+}$. In panel a the results  re shown for $p=0.4$ and $c^{+}=4$. In Supplementary Table 1 and in Supplementary Fig. \ref{figure11S} we provide more information about these datasets.
}
  \label{figure6}
 \end{figure}

In absence of negative triadic interactions, when all regulatory interactions are positive, the dynamics always reaches a stationary point independent of time. In  Fig.~$\ref{figure6}$ a we show a typical time-series for $R^{(t)}$ where it is apparent that $R$ reaches a stationary limit $R^{(t)}=R^{\star}$, where $R^{\star}$ is independent of time. Moreover in Fig.~\ref{figure6}b we also display the dependence of this stationary state with $p$, i.e., $R=R^{\star}(p)$.  
The agreement between theoretical predictions and results of numerical simulations is excellent. Interestingly, the order parameter $R$ displays a discontinuous hybrid phase transition as a function of $p$ showing that positive triadic interactions induce discontinuous hybrid percolation in higher-order networks (see Fig.~$\ref{figure6}$
and the SI  for the analytical derivation of this result). 

 {  
   In order to exclude that  the observed chaotic behavior of triadic percolation is an artefact of the particular choice of the dynamics,
   we consider also a version of the model with time-delayed regulatory interactions, where
   each regulatory link  is assigned a time delay $\tau$
   and
   Step 2 of triadic percolation %with
   is replaced by
   \begin{itemize}
   \item[Step 2$^{\prime}$] Given the set of all active nodes obtained in Step 1,  each structural link is deactivated:
     \begin{itemize}
     \item[(a)] { if none of its positive regulators is active at time at $t -\tau$;
     \item[(b)] if at least one of its negative regulators  is active at time $t -\tau$;}
     \item[(c)] if the structural link is not deactivated according the conditions (a) and (b), it can still be deactivated by stochastic events which occur with probability $q = 1-p$.
       \end{itemize}
\end{itemize}

We consider two models of  triadic percolation  with time delay which depend on the choice of the probability distribution for time delays of regulatory links (see the illustration of the models in Figure $\ref{fig:delay}$):
\begin{itemize}
\item{[Model 1]} each structural link is regulated by regulatory links associated to the same time delay $\tau$, with the time delay $\tau$ being drawn from the distribution $\tilde{p}(\tau)$;
\item{[Model 2]} each regulatory link is associated to a time delay drawn independently from the distribution $\tilde{p}(\tau)$.
\end{itemize}

Note that both models reduce to triadic percolation without delays when $\tilde{p}(\tau)=\delta_{\tau,1}$ where $\delta_{x,y}$ indicates the Kronecker delta.
Interestingly both models lead to a route to chaos also in presence of a non-trivial distribution of time delays,  although the universality class might be different from the one of  the logistic map (see Figure $\ref{fig:delay}$).
This finding demonstrates that the route to chaos observed in triadic percolation is a robust feature of the triadic-percolation model.
Finally we note that triadic percolation might be suitably generalized also to node percolation leading also in this case to a route to chaos for the order parameter $R$ (see SI  and Supplementary Fig. 13 for details about this generalization of triadic percolation).
\begin{figure}[!tbh]
  \includegraphics[width=\linewidth]{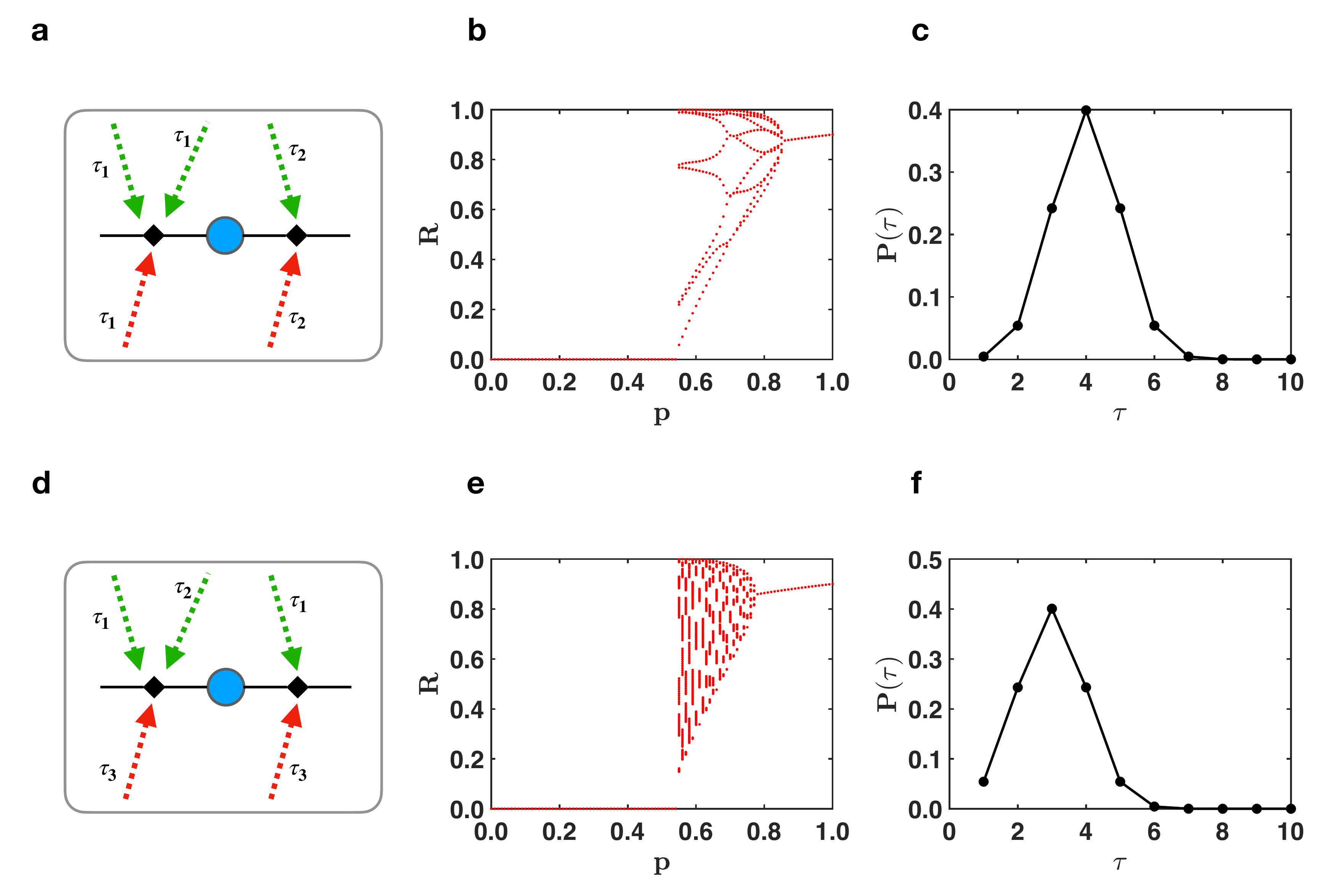}
  \caption{{\bf Triadic percolation with time delays.}
    {   Model 1 and Model 2 of triadic percolation with delays are illustrated in panel (a) and (d) respectively. The corresponding phase diagram for structural Poisson network are shown in panel (b) for Model 1 and in panel (d) for Model 2. The orbit diagrams in panels (b) and (e) are obtained for the
      $\tilde{p}(\tau)$
      distribution of delays shown in panel (c) and (f) respectively. The two orbit diagrams are obtained from the same structural and regulatory network. The structural network is a Poisson network with an average degree $c=50$ and the regulator network has  Poisson degree distribution with average degree $c_{+}=10$ and $c_{-}=3.3$. 
}}
  \label{fig:delay}
\end{figure}
}

\section{Discussion}
A combination of positive and negative interactions is known to affect statistical mechanics problems in non-trivial ways\cite{mezard1987spin,motter2018antagonistic}.  For instance the introduction of signed interactions in the Ising model change dramatically the phase diagram of the model and gives rise to spin glasses with a complex free-energy landscape which display a very different structure of equilibrium configurations with respect to the Ising model. 
Here we  combine the theory of percolation with the theory of  dynamical systems and we show that positive and negative regulatory triadic interactions can turn percolation into a fully fledged dynamical process where the order parameter undergoes a period doubling and a route to chaos.
This implies that, although the underlying structural network remains the same, links and nodes can be activated and deactivated in time leading to a very non-trivial dynamics of the giant component of the network which can ``blink" among few possible connectivity configurations or change in time in a chaotic way. This implies that triadic percolation is radically different from standard percolation in which   the activity of the links is not dynamically regulated and for each value of $p$ characterizing the probability that links are active, the order parameter takes only a single value.
This significant effect of triadic interaction on percolation is captured by the striking difference between the phase diagram of triadic percolation and the phase diagram of standard percolation. While standard percolation leads to a second-order phase transition, the phase diagram of triadic percolation, as long as negative regulatory interactions are included,  becomes an orbit diagram.
In absence of negative regulatory interactions, triadic percolation has an order parameter that  always reaches a steady state and the phase diagram displays  a discontinuous hybrid phase transition.
Our conclusions are based on a general theory giving very accurate predictions although being of a mean-field nature and on extensive simulations performed on synthetic as well as real-world network topologies.

The model can be modified in different ways to address the needs for specific real systems. For instance the approach  can be applied to other generalized network structures such as  hypergraphs and multiplex networks. Moreover the regulatory rules adopted can be modified. Finally, the approach can be extended to situations in which the nodes sustain a more complex dynamics.
 
These results radically change our understanding of percolation and can be used to shed light on real systems in which the functional connectivity of the network is strongly dependent on time as in neuronal and brain networks and in climate. A particularly promising future direction is to apply this theoretical framework to modelling extreme rainfall events. This could lead to a substantial improvement of their forecasting.

\subsection{Data availability.}

All the datasets used in this study are available on the public repository Ref.\cite{Repository_networks}.

\subsection{Code availability.}

{  The codes used in this study are available at the GitHub repository Triadic percolation, https://zenodo.org/record/7651480 {DOI: 10.5281/zenodo.7651480}.}
%\bibliography{references}

\begin{addendum}

 \item[Acknowledgments] G.B. and H.S. thank Franco Vivaldi for interesting discussions.  This research utilized Queen Mary’s Apocrita HPC facility, supported by QMUL Research-IT.
http://doi.org/10.5281/zenodo.438045. G.B. acknowledges support from the Turing-Roche North Star partnership and the Royal Society (IEC\textbackslash NSFC\textbackslash191147. H.S. acknowledges support by  the Chinese Scholarship Council. F.R. acknowledges support by the Air Force Office of Scientific Research (FA9550-21-1-0446) and the Army Research Office (W911NF-21-1-0194). The funders had no role in study design, data collection and analysis, decision to publish, or  any opinions, findings, and conclusions or recommendations expressed in the manuscript. J.K. has been supported by the Alexander von Humboldt Polish Honorary Research Scholarship 2020 of the Fundation for Polish Science.
 
 \item[Author contributions] 
G.B. conceived the project; H.S. F.R. J.K. and G.B. designed the project and wrote the manuscript; H.S., F.R. and G.B. performed the analytical  and the numerical calculations.
 
 \item[Competing interests] The authors declare that they have no competing interests. The funders had no role in study design, data collection and analysis, decision to publish, or preparation of the manuscript.

 \item[Correspondence] Correspondence and requests for materials
should be addressed to ginestra.bianconi@gmail.com
\end{addendum}

%\bibliographystyle{apsrev4-1}
%\bibliography{references}

\clearpage
%%%%%%%%%%%%%%%%%%%%%%%%%%%%%%%%%%%%%%%%%%%%%%%%%%%%%%%%

\begin{center}
	{\Large{\textbf{Supplementary Information \\ ``The dynamic nature of percolation on networks with \\ triadic interactions"}}}\\ \bigskip
	\large{H. Sun, F. Radicchi, J. Kurths and G. Bianconi}
\end{center}	
	\setcounter{section}{0}
	\setcounter{table}{0}
	\renewcommand{\thetable}{S\arabic{table}}
\setcounter{equation}{0}
\renewcommand{\theequation}{S\arabic{equation}}
\setcounter{figure}{0}
\renewcommand{\thefigure}{S\arabic{figure}}

\section{Triadic percolation for correlated structural and regulatory networks}
\subsection{General theoretical framework-}
In this section we extend the theoretical approach described in the main text in order to treat also the case in which the structural degree $k$ of a node can be correlated with its regulatory degrees $\kappa^{+}$ and $\kappa^{-}$.

In this case the random higher-order network with triadic interactions is characterized by a joint degree distribution $\tilde{P}(k,\kappa^{+},\kappa^{-})$ and by the degree distributions $\hat{P}_{\pm}(\hat\kappa^{\pm})$.
The distribution $\tilde{P}(k,\kappa^{+},\kappa^{-})$ indicates the probability that a random node has structural degree $k$ and  regulatory degrees $\kappa^{+}$ and $\kappa^{-}$. The distributions $\hat{P}_{\pm}(\hat\kappa^{\pm})$ indicates the probability that a random link has $\hat{\kappa}^{+}$ or $\hat{\kappa}^-$ regulatory interactions, respectively.

Let us define $S^{(t)}$ as the probability that a node at the endpoint of a  random  structural link of network ${\mathcal {A}}$  is in the giant component (GC) at time $t$.
Let us define $\hat{S}^{(t)\pm}$ as the probability that a node regulating (positively $+$ or negatively $-$) a  random structural link  is in the GC at time $t$.
Let us define with $p_L^{(t)}$ the probability that a random structural link is {active} at time $t$.
By putting $p_L^{(0)}=p_0$ indicating the probability that structural links are active at time $t=0$, we have that for all $t>0$, as long as the network is locally tree-like, $S^{(t)}$, $\hat{S}^{(t)\pm}$ and $p_L^{(t)}$ are updated as 
\bea
S^{(t)}&=&1-G_1\left(1-S^{(t)} p_L^{(t-1)}\right), \nonumber \\
\hat{S}^{(t)\pm}&=&1-\mathcal{G}^{\pm}(1-S^{(t)} p_L^{(t-1)}), \nonumber \\
p_L^{(t)}&=& p  G_0^{-}(1-\hat{S}^{(t)-}) \left[1-G_0^{+}(1-\hat{S}^{(t)+})\right],
\label{reg_percolationS}
\eea
where
\bea
G_1(x)&=&\sum_{k,\kappa^{+},\kappa^{-}}\tilde{P}(k,\kappa^{+},\kappa^{-})\frac{k}{\avg{k}}x^{k-1},\nonumber \\
{G_0}^{\pm}(x)&=&\sum_{{\kappa}_{\pm}}\hat{P}_{\pm}({\hat{\kappa}}^{\pm})x^{{\hat{\kappa}}^{\pm}},\nonumber \\
\mathcal{G^{\pm}}(x)&=&\sum_{k,\kappa^{+},\kappa^{-}}\tilde{P}(k,\kappa^{+},\kappa^{-})\frac{\kappa^{\pm}}{\avg{\kappa^{\pm}}}x^{k}.
\label{Gen_general}
\eea
The probability that a node is in the GC is given by
\bea
R^{(t)}=1-G_0\left(1-S^{(t)} p_L^{(t-1)}\right),
\eea
where
\bea
G_0(x)=\sum_{k, \kappa^{+},\kappa^{-}}\tilde{P}(k,\kappa^{+},\kappa^{-})x^k.
\eea

\subsection{The stationary solution and the onset of its instability-}
The equations for  triadic percolation can be formally written as a map \cite{strogatz2018nonlinear_SI}:
\bea
\hat{S}^{(t)\pm}=f^{\pm}(p_L^{(t-1)}),\quad p_L^{(t)}=g_p(\hat{S}^{(t),+},\hat{S}^{(t),-}),
\eea
whose stationary fixed point  $\hat{S}^{\star\pm},p_L^{\star}$ satisfies
\bea
\hat{S}^{\star\pm}=f^{\pm}(g_p(\hat{S}^{\star,+},\hat{S}^{\star,-})),
\eea
or equivalently 
\bea
p_L^{\star}=g_p(f^{+}({p}_L^{\star}),f^-({p}_L^{\star})).
\eea
The stationary solution becomes unstable when 
\bea
\left|J\right|=1,
\eea
where \bea
J=\left.\frac{\mathrm{d} g_p(f^{+}({p}_L),f^-({p}_L))}{\mathrm{d} p_L}\right|_{p_L=p_L^{\star}}.
\eea
As we will show in the next paragraphs, there are two major types of instability. The first type of instability is observed when $J=1$ and leads to discontinuous hybrid transitions. This type of instability is  observed for instance for triadic percolation in absence of negative interactions. 
The second type of instability is achieved instead when $J=-1$ and this leads typically to the onset of period-$2$ oscillations of the order parameter $R^{(t)}$ of percolation.

Note that the stability of the periodic oscillations of the order parameter can be studied in an analogous way by investigating the stability of the map iterated for a number of times equal to the period of the oscillation under study. However, we leave this analysis to later works.\\

\subsection{Limiting case of uncorrelated structural and regulatory degrees of the nodes-}

If we assume that the structural and regulatory degrees of the nodes are uncorrelated, we can then write
\bea
\tilde{P}(k,\kappa^{+},\kappa^{-})=\pi(k)P(\kappa^{+},\kappa^{-}).
\eea
 The equations (\ref{reg_percolationS}) do simplify as $\mathcal{G^{+}}(x)=\mathcal{G^{-}}(x)$ and the phase diagram is independent of the degree distribution $P(\kappa^{+},\kappa^{-})$.
 Therefore in this limit we recover Eq. (2) of the main text that we repeat here for completeness, \ie,
 \bea
S^{(t)}&=&1-G_1\left(1-S^{(t)} p_L^{(t-1)}\right),\label{reg_percolationS2_a} \\
{R}^{(t)}&=&1-G_0\left(1-S^{(t)} p_L^{(t-1)}\right),\label{reg_percolationS2_b} \\
p_L^{(t)}&=& p  G_0^{-}(1-{R}^{(t)}) \left[1-G_0^{+}\left(1-{R}^{(t)}\right)\right].
\label{reg_percolationS2_c}
\eea
Here, we have used the simplified definition of the generating functions given by 
\bea
G_0(x)&=&\sum_{k}\pi(k)x^k, \nonumber \\
G_1(x)&=&\sum_{k} \pi(k)\frac{k}{\avg{k}}x^{k-1},\nonumber \\
{G_0}^{\pm}(x)&=&\sum_{{\kappa}_{\pm}}\hat{P}_{\pm}({\hat{\kappa}}^{\pm})x^{{\hat{\kappa}}^{\pm}}.
\label{gen_uncorrS}
\eea
As noted in the main text,  Eq.~(\ref{reg_percolationS2_a})-(\ref{reg_percolationS2_c}) for the percolation model regulated by triadic interactions can be formally written as the map \cite{strogatz2018nonlinear_SI}
\bea
R^{(t)}=f\left(p_L^{(t-1)}\right),\quad 
p_L^{(t)}=g_p\left(R^{(t)}\right),
\label{eq:cobweb_bS}
\eea
 or combining these two equations as the map 
 \bea
 R^{(t)}=h_p\left(R^{(t-1)}\right)=f\left(g_p\left(R^{(t-1)}\right)\right).
 \eea
 The stationary solution $R^{(t)}=R^{\star}$ of this map obeys the equation 
 \bea
 R^{\star}=h_p(R^{\star}).
 \label{statS}
\eea 

This stationary solution becomes unstable as soon as 
\bea|J|=1,\eea 
where 
\bea 
J=h_p^{\prime}(R^{\star})=\left.\frac{df}{dp_L}\right|_{p_L=p_L^{\star}}\left.\frac{dg_p}{dR}\right|_{R=R^{\star}}.
\label{Rd}
\eea 
Interestingly, while the value $J=1$ indicates a discontinuous and hybrid transition, the value $J=-1$ indicates the onset of period-2 oscillations. 
%\end{document}

Let us here show that the discontinuous transition observed for $J=1$ is actually hybrid.
To this end we indicate with $p_c$ the value of $p$ for which $J=1$ is satisfied and  we consider small variations $\delta p=p-p_c\ll 1$. We indicate the corresponding change in the stationary solution $R^{\star}$ with $\delta R=R^{\star}(p)-R^{\star}(p_c)\ll 1$.
Since both $R^{\star}(p)$ and $R^{\star}(p_c)=R_c$ satisfy the stationary Eq.~(\ref{statS}), assuming without loss of generality that $h_p(R^{\star})$ is twice differentiable at $R^{\star}(p_c)=R_c>0$  we can expand this latter equation in $\delta p$ and $\delta R^{\star}$ obtaining
\bea
\delta R=h_{p_c}^{\prime}(R_c)\delta R+\frac{1}{2}h_{p_c}^{\prime\prime}(R_c)(\delta R)^2+\frac{\partial h_{p_c}(R_c)}{\partial p}\delta p.
\eea
Since $h_{p_c}^{\prime}(R_c)=1$ this equation reduces to 
\bea
\frac{1}{2}h_{p_c}^{\prime\prime}(R_c)(\delta R)^2+\frac{\partial h_{p_c}(R_c)}{\partial p}\delta p=0,
\eea
from which it is immediate to derive the scaling $\delta R\propto (\delta p)^{1/2}$ as long as $h_{p_c}^{\prime\prime}(R_c)$ and ${\partial h_{p_c}(R_c)}/{\partial p}$ have finite values and opposite sign. Therefore we have shown that  
\bea
R^{\star}(p)-R_c\propto (p-p_c)^{1/2},
\eea
which establishes that the discontinuous transition is hybrid.

 In order  to provide evidence and motivation of the general  result that $J=-1$ indicates the onset of period two oscillations in the next section, we will show a  concrete example. For the moment, let us observe that the explicit expression of $J$ given by Eq.~(\ref{Rd}) implies that  since $df/dp_L\geq 0$ the onset of the period-2 oscillations can take place only if $g_p(R)$ has a negative slope. As a consequence of this we conclude that  period-2 oscillations can occur only if negative regulatory interactions are present. 
 
\begin{figure*}[tbh!]
  \includegraphics[width=\textwidth]{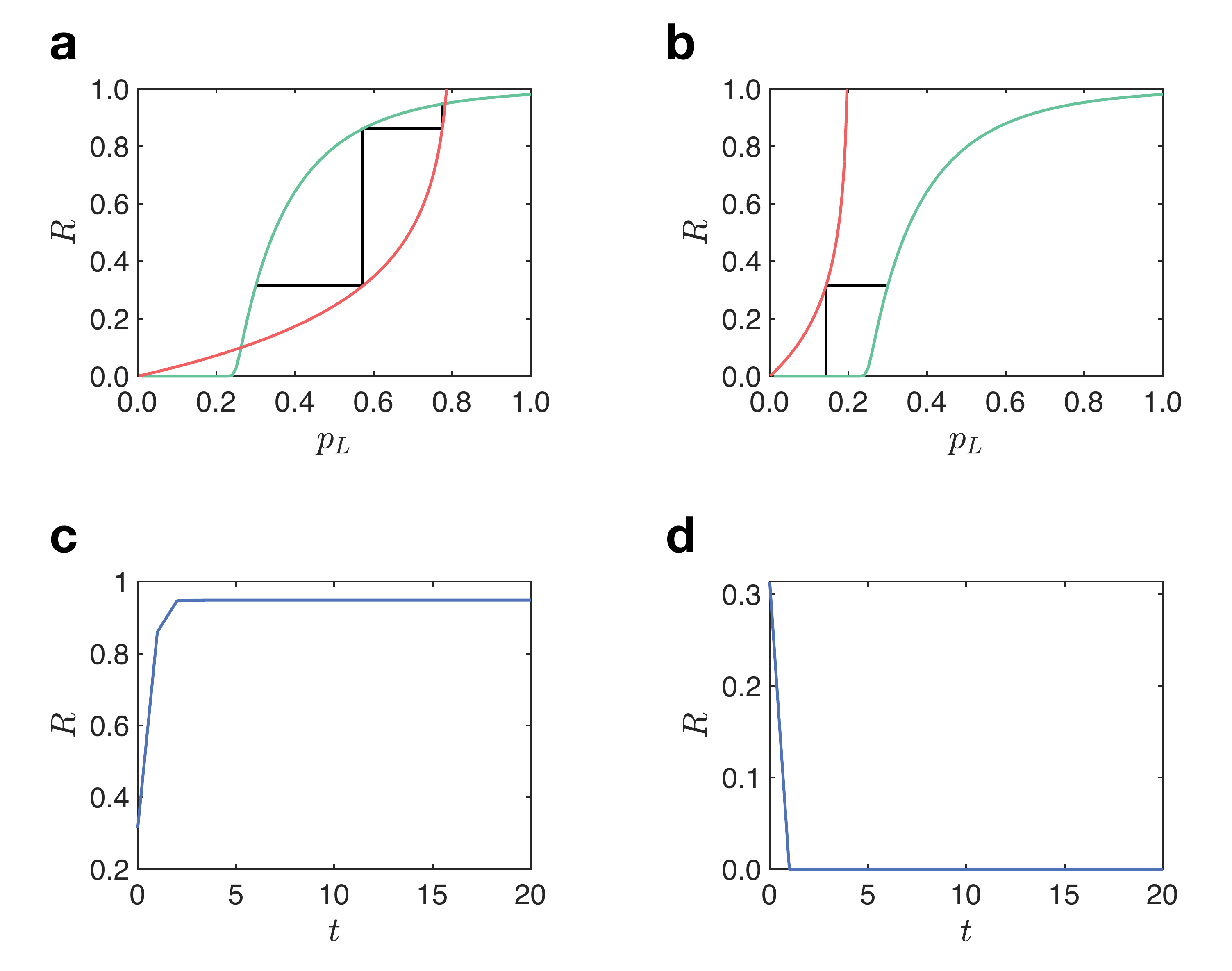}
  \caption{  {The theoretical cobweb plot (panel (a), (b)) and the corresponding  dependence of the order parameter $R$ on time  $t$ (panel (c), (d)) is shown when regulatory interactions are exclusively positive. The structural network is a Poisson network with average degree $c = 4$, and Poisson distribution $\hat{P}_{\pm}(\hat{\kappa}_{\pm})$ with average degrees $c^+ = 4$ and $c^- = 0$ respectively. In panel (a) and (c)  $p=0.8>p_c$; in panel (b) and (d)  $p=0.2 < p_c$. The results are obtained with an initial condition $p_L^{(0)}=0.3$.}}
  \label{figure1S}
  \end{figure*}

\begin{figure*}[tbh!]
  \includegraphics[width=\textwidth]{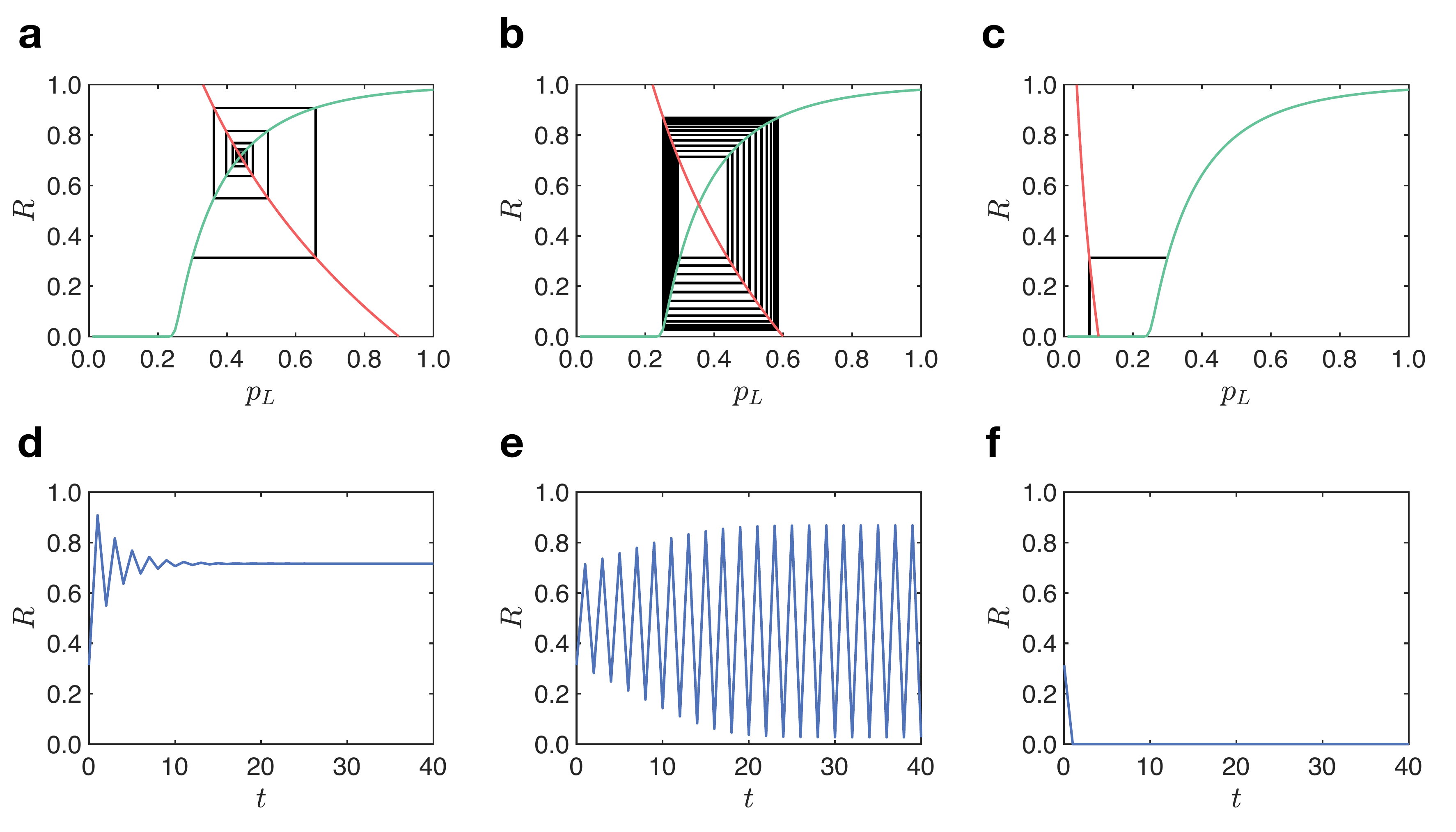}
  \caption{  {The theoretical cobweb plot (panel (a), (b), (c)) and the corresponding dependence of the order parameter $R$ on time  $t$ (panel (d), (e), (f)) is shown when regulatory interactions are exclusively negative. The structural network is a Poisson network with average degree $c = 4$, and Poisson distribution $\hat{P}_{\pm}(\hat{\kappa}_{\pm})$ with average degrees $c^+ = 4$ and $c^+ = \infty$ respectively. In panel (a) and (d)  $p=0.9$; in panel (b) and (e)  $p=0.6$; in panel (c) and (f) $p=0.1$. The results are obtained with an initial condition $p_L^{(0)}=0.3$.}}
  \label{figure2S}
  \end{figure*}

 In Fig.~\ref{figure1S} we show the cobweb when only positive regulatory interactions are present, \ie, when Eq.~(\ref{reg_percolationS2_c}) is substituted by 
\bea
p_L^{(t)}&=& p  \left[1-G_0^{+}\left(1-{R}^{(t)}\right)\right].
\eea 
 In Fig.~$\ref{figure2S}$ we show the cobweb when only the negative regulatory interactions are present and positive interactions do not play a role, \ie, when Eq.~(\ref{reg_percolationS2_c}) is substituted by 
\bea
p_L^{(t)}&=& p  G_0^{-}(1-{R}^{(t)}).
\eea 
 For examples of the cobweb when both positive and negative interactions are present and relevant see Fig. $\ref{figure3}$ of the main text.

\section{Triadic percolation for uncorrelated Poisson structural network}
\subsection{Triadic percolation on uncorrelated Poisson structural networks}
In this section we investigate the instability of the stationary solution in the case of a Poisson structural network of average degree $c$  in which the structural and the regulatory degrees of the nodes are uncorrelated, \ie, 
%in which   
\bea
\tilde{P}(k,\kappa^{+},\kappa^{-})=\pi(k)P(\kappa^{+},\kappa^{-}),
\eea
with 
\bea
\pi(k)&=&\frac{1}{k!}c^k e^{-c}.
\eea
Additionally we assume that $\hat{\kappa}^{+}$ and $\hat{\kappa}^{-}$ are drawn from Poisson distributions with average degree $c^{+}$ and $c^{-}$ respectively, \ie,
\bea
\hat{P}_{\pm}(\hat{\kappa}^{\pm})&=&\frac{1}{\hat{\kappa}^{\pm}!} (c^{\pm})^{\hat{\kappa}^{\pm}} e^{-c^{\pm}}.
\eea
Eq.~(\ref{reg_percolationS2_b}) and Eq.~(\ref{reg_percolationS2_c}) for the triadic percolation reduces to 
\bea\label{eq:poisson}
 R^{(t)}& = &1-e^{-c p_L^{(t-1)}R^{(t)}}\\ \nonumber
p_L^{(t)}&=&p \left(1-e^{-c^+ R^{(t)}}\right)e^{-c^-R^{(t)}}. 
\eea
The first equation can be expressed as a map between $p_L^{(t-1)}$ and $R^{(t)}$, while the second equation can be expressed as a map between $R^{(t)}$ and $p_L^{(t)}$, \ie,
\bea
R^{(t)}=f\left(p_L^{(t-1)}\right), \quad p_L^{(t)}=g_p\left(R^{(t)}\right).
\eea
Both equations can be combined in the single map 
\bea 
R^{(t)}=h_p\left(R^{(t-1)}\right)=f\left(g_p\left(R^{(t-1)}\right)\right).
\eea

\subsection{Onset of the instability of the stationary solutions-}
Triadic percolation on a structural Poisson network and a Poisson regulatory network  with average degrees $c^+$ and $c^-$ for positive and negative regulatory interactions,   admits a stationary steady state  when Eq. (\ref{eq:poisson}) have the solution $R^{(t)}=R^{\star},  p_L^{(t)}=p_L^{\star}$, where $R^{\star}$ and $p_L^{\star}$ satisfy
\bea\label{eq:jacobian1}
R^{\star} &=&  1-e^{-c p_L^{\star} R^{\star}},\nonumber\\ 
p_L^{\star} &=& p \left(1-e^{-c^+R^{\star}}\right)e^{-c^-R^{\star}}.
\eea
These equations can be expressed as a single equation 
\bea
R^{\star}=h_p(R^{\star})=f(g_p(R^{\star})),
\eea
where the functions $h_p(R)$, $f(p_L)$ and $g_p(R)$ have the same definition as in the previous paragraph.
The stationary solution becomes unstable for  
\bea
|J|=|h_p^{\prime}(R^{\star})|=\left\lvert \frac{d f(g_p(R^{\star}))}{d R^{\star}}\right\rvert=\left\lvert f^{\prime}(p_L^{\star})g_p^{\prime}(R^{\star})\right\rvert=1, \label{eq:jacobian2}
\eea
where $f^{\prime}$ and $g^{\prime}$ are given by 
\bea
f^{\prime}(p_L^{\star})&=&-\frac{c R^{\star}}{c p_L^{\star} - e^{c p_L^{\star} R^{\star}}},\nonumber \\
g_p^{\prime}(R^{\star})&=&p(c^{+}+c^-)e^{-(c^{-}+c^+)R^{\star}}-c^-pe^{-c^{-}R^{\star}}.
\eea
Solving Eq.~(\ref{eq:jacobian1}) and Eq.~(\ref{eq:jacobian2}) numerically when $J=1$ we find the critical manifold  of discontinuous hybrid transitions and when $J=-1$ we find the manifold  for the onset of period-2 oscillations of the order parameter $R^{(t)}$.
\begin{figure*}
\begin{center}
  \includegraphics[width=0.7\textwidth]{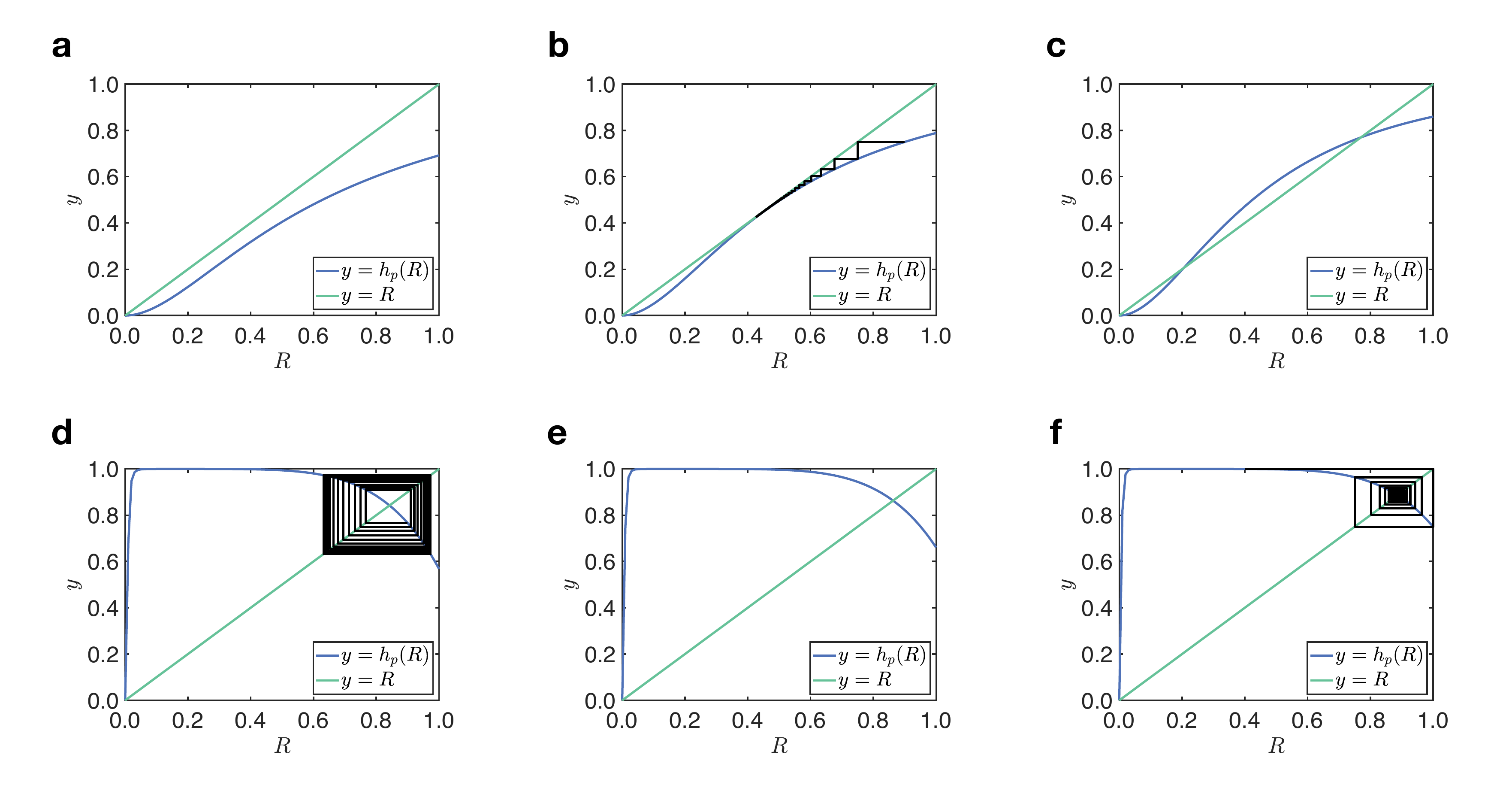}
  \end{center}
  \caption{The figures show the two different modalities for the onset of the instability of the stable solution of the iterative map $R^{(t)}=h_p(R^{(t-1)})$ for a Poisson network with triadic interactions corresponding to the crossing of the curves $y=h_p(R)$ and $y=R$. In panels (a), (b),and (c) we show the emergence of the discontinuous transition at $p=0.392$ (panel (b)) on a  Poisson network with average degree $c=4$, and Poisson distribution s $\hat{P}^{\pm}(\hat{\kappa}^{\pm})$ with average degrees $c^+=4$ and $c^-=0$ respectively. Panels (a)  and (c) show the functions $y=h_p(R)$ and $y=R$ for $p=0.30$ (below the transition) and $p=0.50$ (above the transition).  
  Note that in panel (b) the function $y=h(R)$ and the function $y=R$ are tangent to each other at their non-trivial intersection indicating that the non-trivial solution disappears as soon as $p<0.392$. In panel (d), (e), (f) we show the emergence of 2-cycle  at $p=0.665$ (panel (e)) for a Poisson network with average degree $c=30$, and Poisson distributions  $\hat{P}^{\pm}(\hat{\kappa}^{\pm})$ with average degree $c^+=10$ and $c^-=2.5$ respectively. Panels (d) and (f) show the functions $y=h(R)$ and $y=R$ for $p=0.60$ (below the transition) and  $p=0.8$ (above the transition) respectively. Note that in panel (e) the function $y=h_p(R)$ displays a derivative $-1$ leading to the emergence of the 2-limit cycle observed for  $p\leq 0.665.$ The relative cobweb are shown only for panels (b), (d) and (f) to improve the readability of the figure.}
  \label{figure3S}
  \end{figure*}
  \begin{figure*}[tbh!]
  \includegraphics[width=\textwidth]{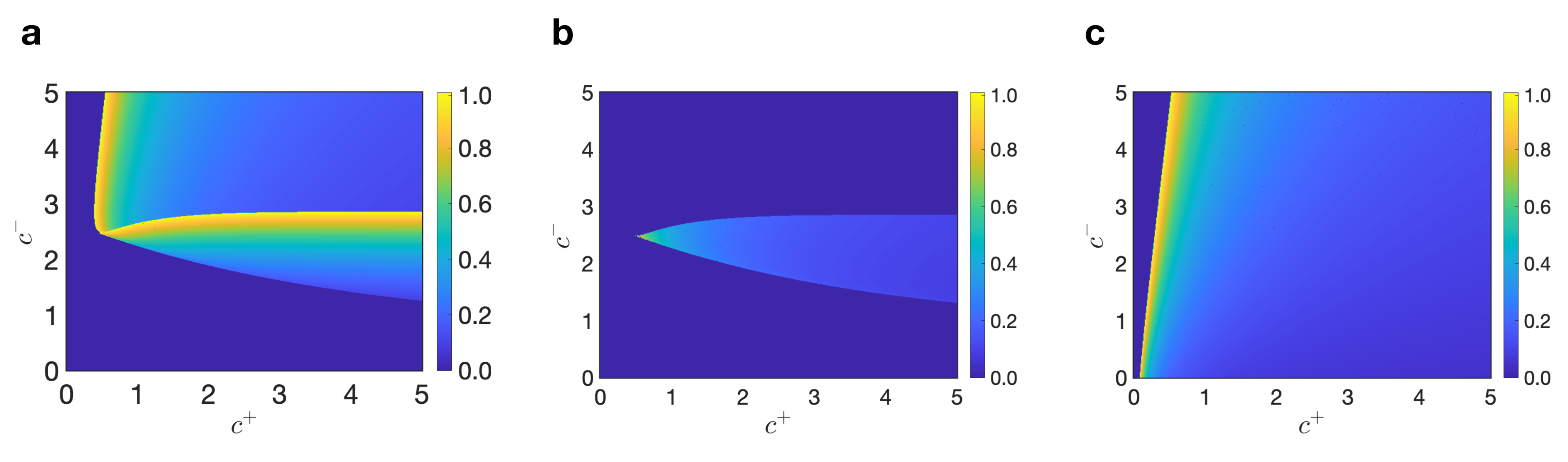}
  \caption{The upper critical point $p_c^u$ (panel a) and the lower critical point $p_c^l$ (panel b)  determining the onset of period-2 oscillations are plotted plane $(c^+,c^-)$.  Panel c represents the critical point $p_c$ at which the discontinuous hybrid transition is observed in the plane $(c^+,c^-)$. In all panels the structural network has Poisson degree distribution with degree $c=30$ the regulatory network has also a Poisson degree distribution  with  $c^+$ and $c^-$ indicating the average positive and negative degree respectively. }
  \label{figure4S}
  \end{figure*}
  
In Fig.~$\ref{figure3S}$ we show graphically the difference between the two types of possible instabilities of the stationary solution $R^{\star}=h(R^{\star})$. When $J=h^{\prime}(R^{\star})=1$ one observes the discontinuous emergence of a non-zero stationary solution $R^{\star}>0$. When $J=h^{\prime}(R^{\star})=-1$ we observe the onset of period-2 oscillations of the order parameter satisfying the map $R^{(t)}=h(R^{(t-1)})$.

In Fig.~$\ref{figure4S}$ we show the obtained critical manifolds for the  onset of period two  oscillations of the order parameters  and for the onset of discontinuous hybrid transitions. Note that in for any given structural and regulatory networks the critical point of the onset of the discontinuous hybrid transition is unique, if such transition exist. However the onset of the period two oscillations can occur for different values of $p$. In Fig.~$\ref{figure4S}$ we plot exclusively the larger and the smaller critical points for the onset of period two oscillations if they exist.

{
\section{Universality class of the route to chaos of triadic percolation}
In the previous sections we have studied triadic percolation in different settings and we have shown that the process can undergo a period doubling transition.
In this section  we demonstrate that triadic percolation undergoes a route to chaos in the universality class of the logistic map as long as the structural and regulatory degrees are uncorrelated and the distributions $P(\hat{\kappa}^{\pm})$ are Poisson.
%For ease of notation we will  demonstrate this is the simple case of a Poisson structural uncorrelated network and uncorrelated  triadic interactions, but this approach can be extended directly to the most general framework of triadic percolation.
\subsection{Logistic map universality class}
Triadic percolation can be captured at the mean-field level by a map \bea
R^{(t)}=h(R^{(t-1)})\eea determining the relative size $R^{(t)}$ of the giant component at time $t$, given the relative size $R^{(t-1)}$ of the giant component at time $t-1$. 
Examples of these maps  obtained from uncorrelated structural Poisson networks are shown in Figure $\ref{fig:map}$.
\begin{figure}[!htb]
  \includegraphics[width=\linewidth]{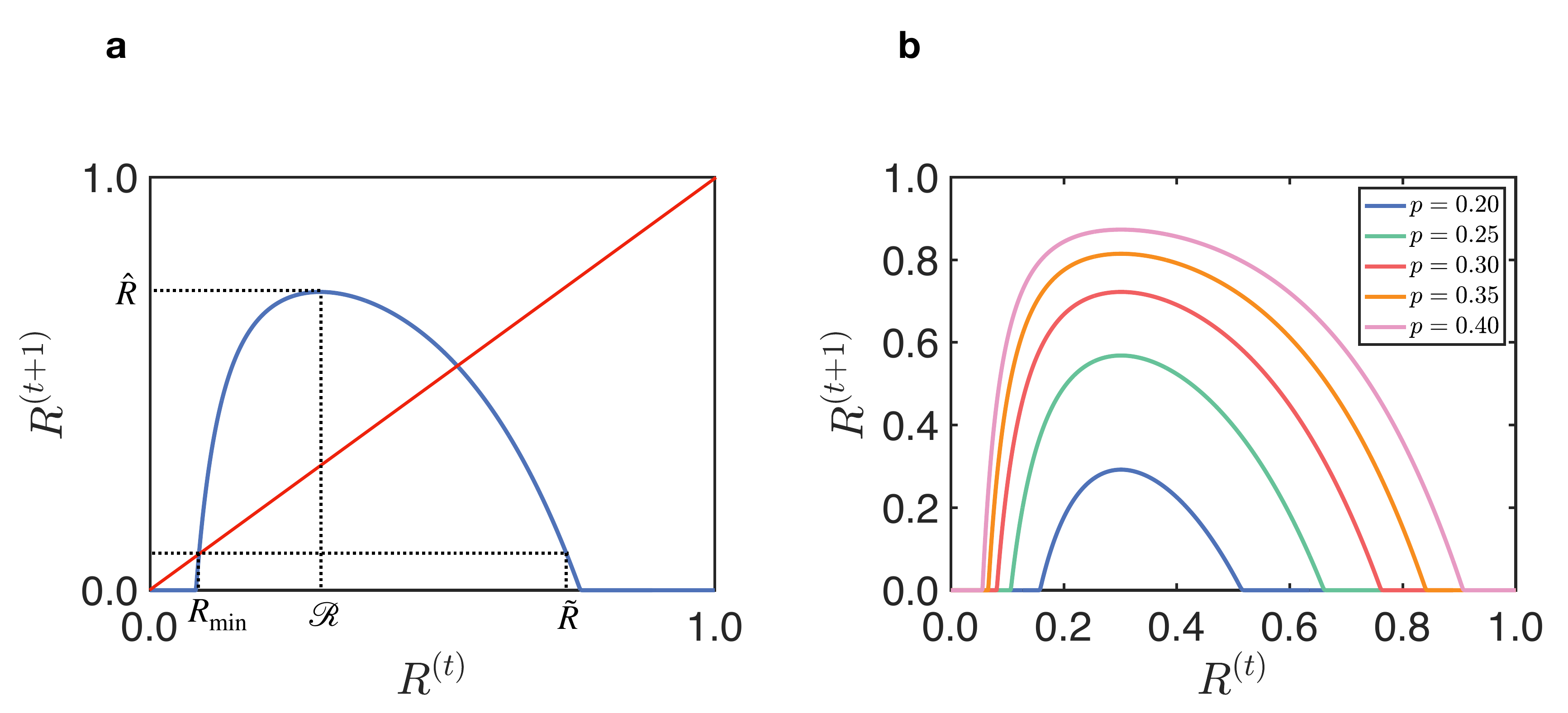}
  \caption{{ {\bf Example of the maps capturing  triadic percolation.}  Panel (a) shows the map  $R^{(t)}=h(R^{(t-1)})$ (in blue) obtained for  $p=0.3$ in the case of a  Poisson structural network and uncorrelated regulatory network with Poisson distributions $P(\hat{\kappa}^{\pm})$.
  The intersections between the red line $R^{(t)}=R^{(t-1)}$ and the map (indicated in blue) determine the fixed points. Here $\underline{R}$ denotes the minimum non-trivial fixed point of the map. The map reaches its maximum $\hat{R}$ for $R=R^{\star}$. We denote with $\bar{R}> \underline{R}$  the point where $h(\bar{R}) = \underline{R}$. Panel (b) displays the map $R^{(t)}=h(R^{t-1})$ for different parameter $p$. 
   In both panels the structural network is a Poisson network with average degree $c=30$ and the distributions $\hat{P}_{\pm}(\hat{\kappa}_{\pm})$ are Poisson with  average degrees $\avg{\hat{\kappa}^+}=c^{+}=1.8$, $\Avg{\hat{\kappa}^-}=c^{-}=2.5$.}}
  \label{fig:map}
\end{figure}
Here we show that this map  is in the universality class of the logistic map.
In order to show that, according to Feigenbaum classic result\cite{feigenbaum1978quantitative_SI}, it is enough to demonstrate that  the function $h(R)$ is unimodal, \ie, has a single maximum at $R=R^{\star}$,
and that close to its maximum, \ie, for $|R-R^{\star}|\ll 1$,
 the function $h(R)$  has a quadratic approximation, with 
\bea
h(R)\simeq h({R}^{\star}) + \frac{1}{2}h^{\prime\prime}({R}^{\star})(R-{R}^{\star})^2.
\eea

To demonstrate this scaling of the function $h(R)$ close to its maximum we provide here the explicit expression of the derivative $dR^{(t)}/dR^{(t-1)}$ in terms of $R^{(t-1)}$ and $R^{(t)}=h(R^{(t-1)})$.

Our starting point will be the formulation of triadic percolation for uncorrelated structural and regulatory degrees of the nodes dictated by the Eqs. (\ref{reg_percolationS2_a}), (\ref{reg_percolationS2_b}) and (\ref{reg_percolationS2_c}), which we rewrite here for completeness,
\bea
S^{(t)}&=&1-G_1\left(1-p_L^{(t-1)}S^{(t)}\right)=F_1\left(p_L^{(t-1)}, S^{(t)}\right),
 \label{reg_percolationS2_ax}\\
R^{(t)}&=&1-G_0\left(1-p_L^{(t-1)}S^{(t)}\right)=F_2\left(p_L^{(t-1)},S^{(t)}\right), \label{reg_percolationS2_bx}\\
p_L^{(t)} &=& p G_0^{-}\left(1-R^{(t)}\right)\left[1-G_0^{+}\left(1-R^{(t)}\right)\right]=F_3\left(R^{(t)}\right),
\label{reg_percolationS2_cx}
\eea
where $G_1(x)$, $G_0(x)$ and $G_0^\pm (x)$ are defined in Eq. (\ref{gen_uncorrS}).
Starting from Eq. (\ref{reg_percolationS2_ax}) and using the chain rule we get
\bea
\frac{d S^{(t)}}{d R^{(t-1)}}=\frac{\partial F_1}{\partial p_L^{(t-1)}}\frac{d p_L^{(t-1)}}{d R^{(t-1)}}+\frac{\partial F_1}{\partial S^{(t)}}\frac{d S^{(t)}}{d R^{(t-1)}}.
\eea
Thus,
\bea
\frac{d S^{(t)}}{d R^{(t-1)}}&=&\frac{\partial F_1}{\partial p_L^{(t-1)}}\frac{d p_L^{(t-1)}}{d R^{(t-1)}}  \left(1-\frac{\partial F_1}{\partial S^{(t)}}\right)^{-1}.
\label{S_dev_ap}
\eea
Similarly we can use the chain rule starting from Eq. (\ref{reg_percolationS2_bx}) to express the derivative ${d R^{(t)}}/{d R^{(t-1)}}$, \ie,
\bea
\frac{d R^{(t)}}{d R^{(t-1)}}=\frac{\partial F_2}{\partial p_L^{(t-1)}}\frac{d p_L^{(t-1)}}{d R^{(t-1)}}+ \frac{\partial F_2}{\partial S^{(t)}}\frac{d S^{(t)}}{d R^{(t-1)}}. 
\eea
Using Eq.(\ref{S_dev_ap}) and the relation 
\bea
\frac{\partial F_2}{\partial p_L^{(t-1)}}\frac{\partial F_1}{\partial S^{(t)}}=\frac{\partial F_1}{\partial p_L^{(t-1)}}\frac{\partial F_2}{\partial S^{(t)}},
\eea
we obtain
\bea
\frac{d R^{(t)}}{d R^{(t-1)}}=\frac{\partial F_2}{\partial p_L^{(t-1)}}\frac{d p_L^{(t-1)}}{d R^{(t-1)}}\left(1-\frac{\partial F_1}{\partial S^{(t)}}\right)^{-1},
\label{dRRx}
\eea
where
\bea
\frac{\partial F_1}{\partial p_L^{(t-1)}} = S^{(t)} \avg{k}G_1\left(1-p_L^{(t-1)}S^{(t)}\right),\quad
\frac{\partial F_1}{\partial S^{(t)}}= p_L^{(t)}\avg{k} G_1\left(1-p_L^{(t-1)}S^{(t)}\right),\nonumber\\
\frac{\partial F_2}{\partial p_L^{(t-1)}}= S^{(t)} G_1^\prime\left(1-p_L^{(t-1)}S^{(t)}\right),\quad
\frac{\partial F_2}{\partial S^{(t)}}= p_L^{(t)} G_1^\prime\left(1-p_L^{(t-1)}S^{(t)}\right), 
\label{px}\eea
and 
\bea
\hspace{-15mm}\frac{d p_L^{(t-1)}}{d R^{(t-1)}} &=& p \left[G_0^{-}(1-R^{(t-1)}) \Avg{\hat{\kappa}^+}G_1^{+}(1-R^{(t-1)})-\avg{\hat{\kappa}^-}G_1^{-}(1-R^{(t-1)}) \left(1-G_0^{+}(1-R^{(t-1)}\right)\right].\nonumber
\eea
Note that here $G_1^{\prime}(x)=\sum_{k}k(k-1)x^{k-2}/\avg{k}$.
From Eq. (\ref{dRRx}) and Eqs. (\ref{px}) it follows that the derivative $dR^{(t)}/dR^{(t-1)}$ vanishes if and only if either $S^{(t)}=0$ or $dp_L^{(t-1)}/dR^{(t-1)}=0$.
Consequently, the maximum of the map is determined by the condition $dp_L^{(t-1)}/dR^{(t-1)}=0$.
Let us now consider the case in which the distributions $P(\hat{\kappa}^{\pm})$ are Poisson with average degree $c^{\pm}$.
In this case $G_0^+(1-R^{(t-1)})=G_1^{+}(1-R^{(t-1)})=\exp(-c^+R^{(t-1)})$ and $G_0^-(1-R^{(t-1)})=G_1^{-}(1-R^{(t-1)})=\exp(-c^-R^{(t-1)})$ and hence
\bea
\frac{\partial p_L^{(t-1)}}{\partial R^{(t-1)}}&=&pe^{-c^-R^{(t-1)}}\left[-c^-+(c^{+}+c^-)e^{-c^+R^{(t-1)}}\right].
\eea
In this case there  is only one singular value $R^{(t-1)}=R^{\star}$ at which $dp_L^{(t-1)}/dR^{(t-1)}=0$ given by 
\bea
R^{\star}=\frac{1}{c^+}\ln\left(\frac{c^++c^-}{c^-}\right).
\eea
It is straightforward to show that 
 \bea
\left.\frac{d^2p_L^{(t-1)}}{d(R^{(t-1)})^2}\right|_{R^{(t-1)}=R^{\star}}=-pe^{-(c^-+c^{+})R^{\star}}(c^++c^-)c^+,
\eea
and,  as long as $\hat{R}=h(R^{\star})>0$,
%from this
it follows immediately  that 
\bea
\left.\frac{d^2R^{(t)}}{d(R^{(t-1)})^2}\right|_{R^{(t-1)}=R^{\star}}=\left.\frac{\partial F_2}{\partial p_L^{(t-1)}}\frac{\partial^2 p_L^{(t-1)}}{\partial (R^{(t-1)})^2}\left(1-\frac{\partial F_1}{\partial S^{(t)}}\right)^{-1}\right|_{R^{(t-1)}=R^{\star}}<0.
\eea
Hence the scaling of the map close to the maximum is quadratic proving that the universality class of triadic percolation is the one of the logistic map as long as the structural and the regulatory degrees are uncorrelated and  $P(\hat{\kappa}^{\pm})$ are Poisson distributions. 

\subsection{Stability of solutions and attractors}

Here we study the stability and the basin of attraction of the trivial solution $R^{(t)}=0$ and the basin of attraction of the non-trivial attractor. In order to do we will investigate the major properties of the map function $R^{(t)}=h(R^{(t-1)})$ and we will make reference to the notation illustrated in Figure $\ref{fig:map}(a)$.
 
For determining  the basin of attraction of the zero solution $R^{(t)}=0$ let us  define 
$\underline{R}$ as the smallest non-zero fixed point
\bea
\underline{R}&=&h(\underline{R}),
\eea
and let us consider $R^{(t-1)}<\underline{R}$.
since $R^{(t)}<R^{(t-1)}$ for all $t$ we derive that any initial condition $R^{(0)}<\underline{R}$ will eventually converge to the zero solution.
Let us consider all the initial conditions $R^{(0)}>\bar{R}$ where $\bar{R}$ is the largest solution of the equation $\underline{R}=h(\bar{R})$.
It is straightforward to see that also all these initial conditions will converge to the zero solution since after the first iteration of the map the problem can be reduced to the previous scenario.

Let us now establish the conditions that will ensure the stability of  the non-trivial attractor.
From the theory of coupled maps \cite{strogatz2018nonlinear_SI}, we are guaranteed that the  non-trivial attractor of the map will be stable as long as the interval $ (\underline{R},\bar{R})$  is mapped into itself or into one of its subsets by the map, \ie, as long as $\hat{R}=h(R^{\star})<\bar{R}$. 
}

\section{Tuning the positive and negative regulatory interactions}
\begin{figure*}[tbh!]
  \includegraphics[width=\textwidth]{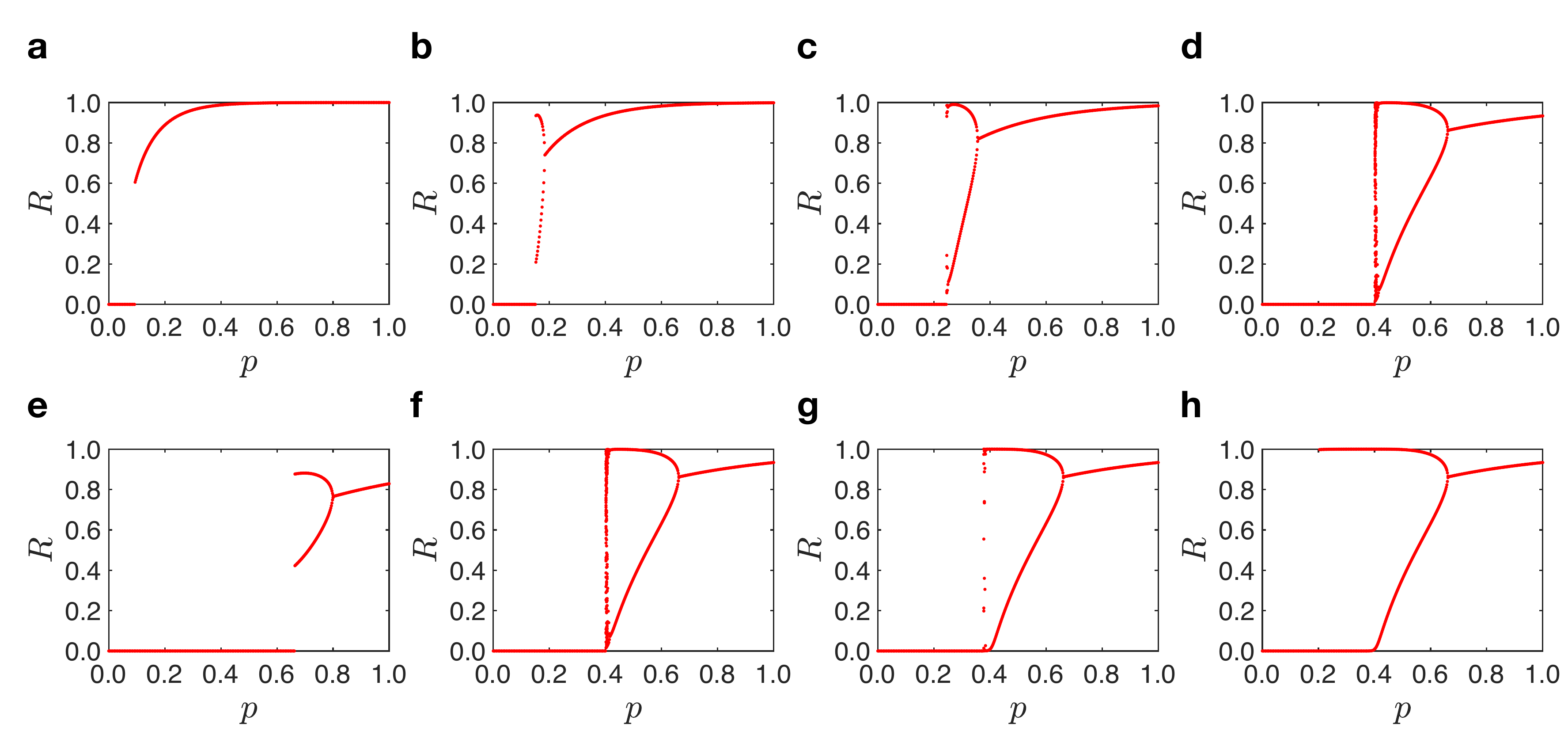}
  \caption{Theoretically obtained orbit diagrams for the Poisson structural network with average degree $c=30$ and uncorrelated structural and regulatory degrees of the nodes. In the first row, $c^+=10$, from the left to the right we increase the $c^-$ that $c^-=1.0$ (a), $c^-=1.5$ (b), $c^-=2.0$ (c), $c^-=2.5$ (d). In the second row, $c^-=2.5$, from the left to the right we increase the $c^+$ that $c^+=1$ (e), $c^+=10$ (f), $c^+=1000$ (g), $c^+=\infty$ (h). For all  panels $c^{\pm}$ indicates the average degree of the Poisson distribution $\hat{P}_{\pm}(\hat{\kappa}^{\pm})$. All figures  are obtained by setting the  initial condition $p_L^{(0)}=0.1$.}
  \label{figure5S}
  \end{figure*}
\begin{figure*}[tbh!]
  \includegraphics[width=\textwidth]{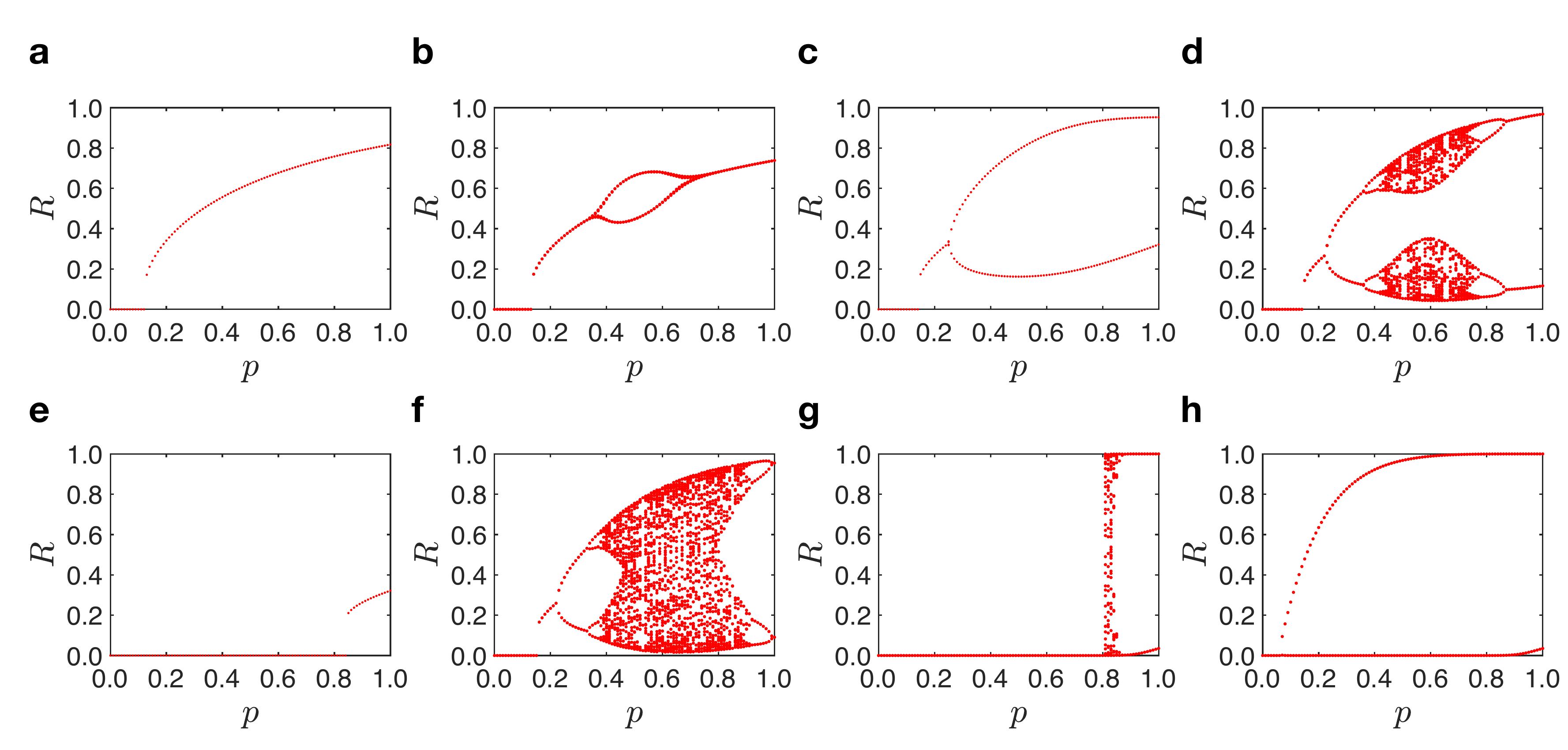}
  \caption{Theoretically obtained orbit diagrams of the  scale-free structural network  with minimum degree $m=4$, power-law exponent $\gamma=2.5$, maximum degree $K=100$ and uncorrelated structural and regulatory degrees of the nodes. In the first row, $c^+=10$, from the left to the right we increase the $c^-$ that $c^-=1.5$ (a), $c^-=1.9$ (b), $c^-=2.3$ (c), $c^-=2.7$ (d). In the second row, $c^-=2.8$, from the left to the right we increase the $c^+$ that $c^+=1$ (e),$c^+=10$ (f), $c^+=1000$ (g), $c^+=\infty$ (h). In all the panels $c^{\pm}$ indicate the average degree of the Poisson distribution $\hat{P}_{\pm}(\hat{\kappa}^{\pm})$. All figures are obtained by setting the initial condition to $p_L^{(0)}=0.1$.}
  \label{figure6S}
  \end{figure*}
  
Triadic percolation  admits two limiting scenarios: the limit $c^-\to 0$ in which the model includes only positive regulatory interactions and is insensitive to negative regulations,  and the limit $c^+\to\infty$ in which the role of positive  regulatory interactions becomes negligible. In fact if $c^-\to 0$ then the condition that none of the negative regulators is active is always satisfied. On the contrary if  $c_+\to \infty$ it becomes sure  that at least one of the infinite positive regulators is active, so the role of positive regulators becomes negligible.
In Fig.~$\ref{figure3S}$ and $\ref{figure4S}$ we investigate the theoretically predicted orbit diagrams of triadic percolation for a Poisson structural network and for a scale-free structural network with structural degree uncorrelated with the regulatory degrees as a function of the average degrees  $c^+$ and $c^-$ of the Poisson distributions $\hat{P}_{\pm}(\hat{\kappa}^{\pm})$.
In absence of negative regulators, \ie, $c^-\to 0$, we observe a discontinuous hybrid transition in both cases (although displaying a smaller discontinuity for the scale-free structural network). 
%In when positive regulatory role is removed 
For $c^+=\infty$, we observe period-2 oscillations in the case of the Poisson structural  network and  a single stable solution in the scale-free case. In both cases, we observe chaos only in presence of both positive and negative regulatory interactions.

  \section{The role of the degree distribution of the structural network}
 In this paragraph we compare the phase diagram of triadic percolation for a scale-free structural network and for a Poisson structural network with the same average degree.

 It is well known that for standard percolation the transition is always continuous and second order with critical indices depending on the second moment of the degree distribution and therefore differing for scale-free networks with a power-law exponent $\gamma\leq 3$ and for Poisson networks. Moreover for the standard bond percolation, scale-free networks  display a zero  percolation threshold in the infinite network limit while  Poisson networks with the same average degree have a finite percolation threshold \cite{dorogovtsev2008critical_SI}. This demonstrates the robustness of scale-free networks under random damage in the framework of the standard percolation theory. Indeed the hubs of scale-free networks, connecting a large set of nodes, keep the network connected also for an extensive entity of the damage of the links. 
 
 On the contrary, for interdependent percolation of multiplex networks  the transition is discontinuous and scale-free networks are more fragile than Poisson networks with the same average degree \cite{buldyrev2010catastrophic_SI}. This phenomenon is   revealed by the  percolation threshold of scale-free networks which is larger than the Poisson network with the same average degree and increasing with the power-law exponent $\gamma$. This phenomenon is due to the fact that hub nodes might be damaged easily if they are interdependent with nodes of small or average degree regardless of the state of their links within their layer.

In absence of negative regulatory interactions,  the present model of triadic percolation displays a discontinuous  transition  as for interdependent networks. However structural  scale-free networks remain more robust than structural Poisson networks. 

\begin{figure*}[tbh!]
  \includegraphics[width=\textwidth]{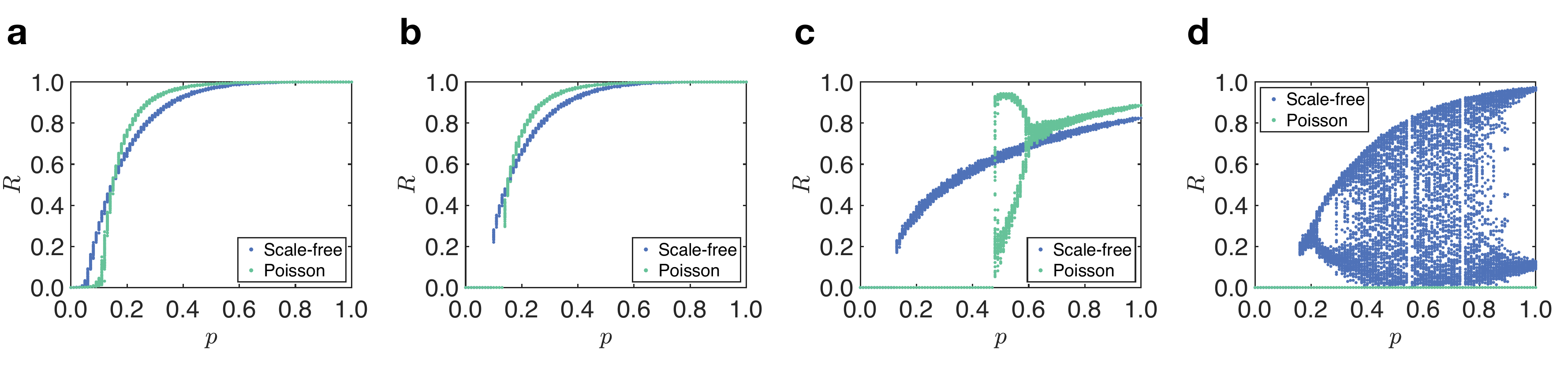}
  \caption{The comparison of orbit diagram between scale-free network and Poisson network with the same averaged degree. The structural scale-free graph has degree exponent $\gamma=2.5$, minimum degree $m=4$ and maximum degree $K=100$. The regulatory network of both scale-free and Poisson structural network has a Poisson distribution with parameters (a) $c^+=\infty$ and $c^-=0$ (standard link percolation); (b) $c^+=10$ and $c^-=0$; (c) $c^+=10$ and $c^-=1.5$ and (d) $c^+=10$ and $c^-=2.8$.}
  \label{figure7S}
  \end{figure*}

In fact  the percolation threshold at which the discontinuous transition occurs is smaller for scale-free networks than for Poisson networks with the same average degree (see Fig.$\ref{figure7S}a$). This is due to the fact that in triadic percolation the regulation  acts directly on the links, and not on the nodes. Therefore the hubs can still play the role of keeping the structural network together also if the percolation transition becomes discontinuous.

When the average degree of the negative regulatory interactions in increased, the comparison of the phase diagram of the structural scale-free networks and the structural Poisson network with the same average degree continues to indicate a larger robustness of the structural scale-free network (see Fig.~$\ref{figure7S}b, c$ and $d$). 

Indeed for larger values of  average degree of negative regulatory interactions  the structural Poisson network displays a period doubling and  chaos of the order parameter while the structural scale-free networks  display a discontinuous phase transition and small fluctuations of the order parameter (see Fig.~$\ref{figure7S}c$). For even  larger values of the average degree of the negative regulatory interactions the structural Poisson network is dismantled for any possible value of $p$ while the structural scale-free networks display period doubling and a route to chaos as a function of $p$ (see Fig.~$\ref{figure7S}d$).

  \section{Comparison between the theory and the Monte Carlo simulations}

\begin{figure*}[tbh!]
  \includegraphics[width=\textwidth]{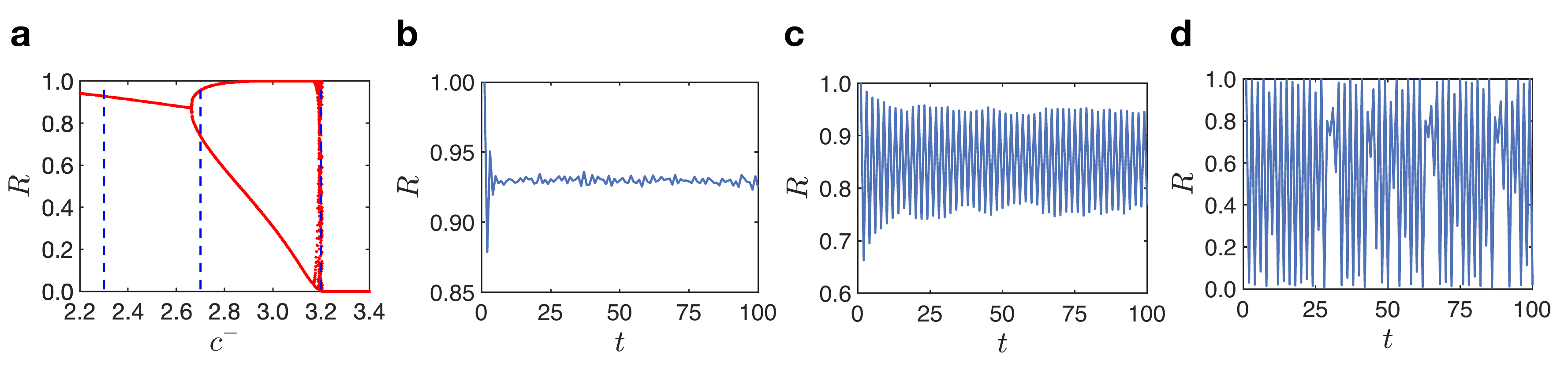}
  \caption{Theoretically obtained orbit diagram at fixed value $p=0.8$ for the  Poisson structural network, with average degree $c=30$,  uncorrelated structural and regulatory degree, Poisson distributed $\hat{\kappa}^{\pm}$ with average $\avg{\hat{\kappa}^{+}}=c^+=10$ and $\avg{\hat{\kappa}^{+}}=c^-$. The three blue lines in panel (a) indicate $c^-=2.3$, $c^-=2.7$ and $c^-=3.2$. The corresponding Monte Carlo simulations on networks of  $N=10^4$  are shown in panel (b),(c) and (d). All figures are obtained with an initial condition $p_L^{(0)}=0.1$.}
  \label{figure8S}
  \end{figure*}

\begin{figure*}[tbh!]
  \includegraphics[width=\textwidth]{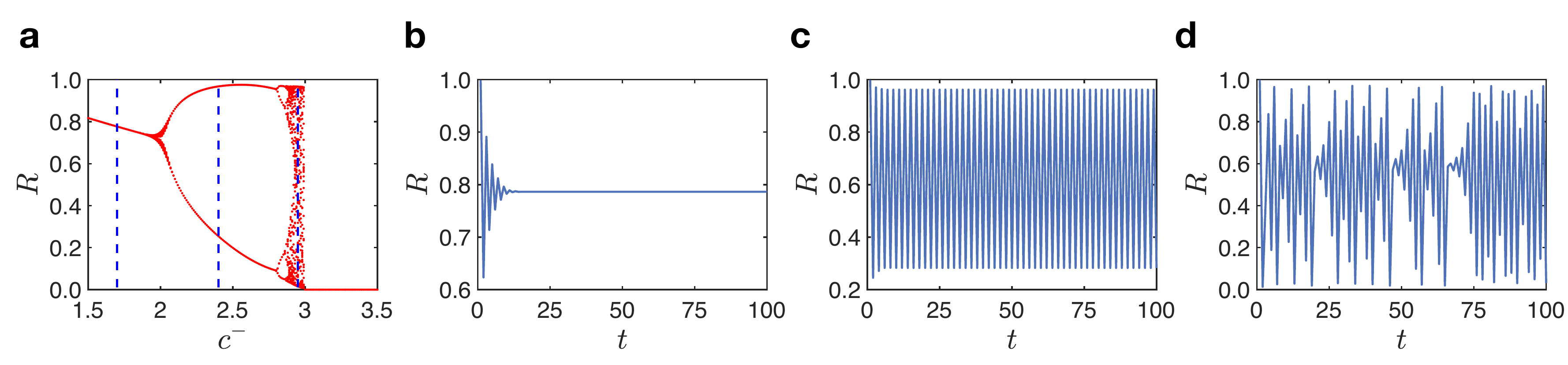}
  \caption{Theoretically obtained orbit diagram at fixed value $p=1$ for the scale-free  structural network, with minimum degree $m=4$, maximum degree $K=100$, power-law exponent $\gamma=2.5$. The nodes have uncorrelated structural and regulatory degrees, and the distribution of $\hat{\kappa}^{\pm}$  is Poisson with average $\avg{\hat{\kappa}^{+}}=c^+=10$ and $\avg{\hat{\kappa}^{+}}=c^-$. The three blue lines in panel (a) indicate $c^-=1.7$, $c^-=2.4$ and $c^-=2.95$. The corresponding Monte Carlo simulations on a network of $N=10^4$ nodes  are shown in panel (b), (c) and (d). All figures are obtained with an initial condition $p_L^{(0)}=0.1$.}
  \label{figure9S}
  \end{figure*}

In the main text, we have studied the period-doubling cascade and the route to chaos of triadic percolation as a function of the probability $p$ that a link is active when all the regulatory interactions are satisfied.
However, here we show that the period-doubling cascade and the route to chaos can also be observed for fixed value of $p$ as a function of the average degree $c^-$ of the Poisson distribution $\hat{P}_-(\hat{\kappa}^-)$ in the case of uncorrelated structural and regulatory degrees of the nodes.
In Fig.~$\ref{figure8S}$ and $\ref{figure9S}$, we show the theoretically obtained orbit diagram  as a function of $c^-$ for a Poisson and for a scale-free network respectively and we compare the theoretical predictions with Monte Carlo simulations of triadic percolation for different values of $c^-$ finding very good agreement.

\begin{figure*}
\begin{center}
   \includegraphics[width=0.6\textwidth]{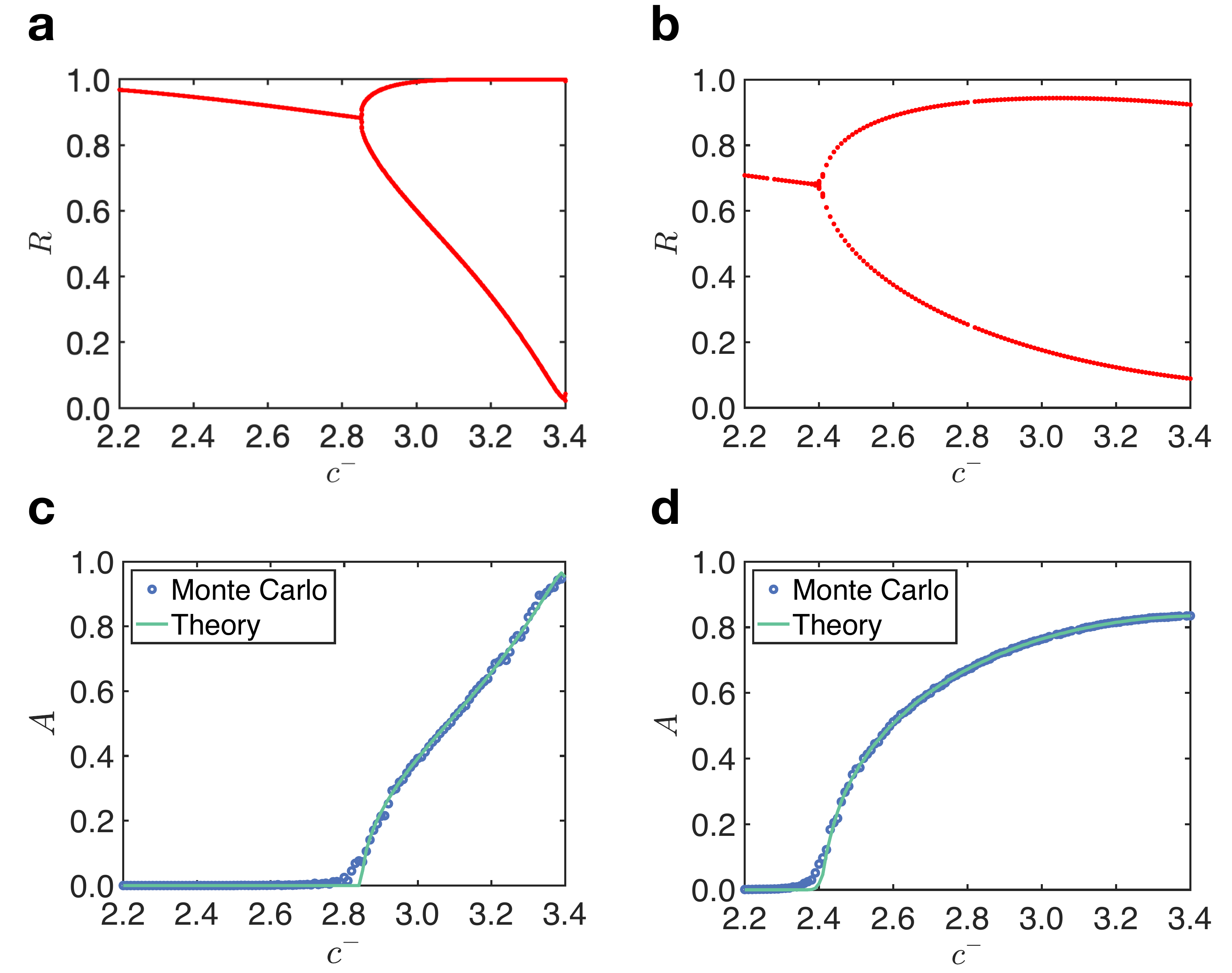}
   \end{center}
  \caption{In panel (a) and (b) we show the theoretically obtained orbit diagram of triadic percolation on a  Poisson network (a) and a scale-free network (b) for $p=1$ as a function of $c^{-}$ indicating the average degree of the Poisson distribution $\hat{P}_{-}(\hat{\kappa}^{-})$. In panel (c) and (d),  we plot the amplitude $A=\Avg{|R^{(t+1)}-R^{(t-1)}|}$ of the period-$2$ oscillations of the order parameter obtained by Monte Carlo simulations (blue circles) of triadic percolation as a function of the average degree $c^-$ (panel (c) refers to the same Poisson network as panel (a) and panel (d) refers to the same scale-free  network as panel (b)). The theoretical prediction for the amplitude $A$ are indicated with a green solid line in panels (c) and (d). 
The Poisson network (panels (a) and (c)) has average structural degree $c=30$ and average degree of the Poisson distribution $\hat{P}_{+}(\hat{\kappa}^{+})$ equal to  $c^+=10$.The scale-free network  (panels (b) and (d)) has minimum degree $m=4$, power-law exponent $\gamma=2.5$ and average degree of the Poisson distribution $\hat{P}_{+}(\hat{\kappa}^{+})$ equal to  $c^+=10$.The Monte Carlo simulations are conducted on networks of $N=10^4$ nodes; the amplitude $A$ is obtained by averaging over $10$ network realizations. }
  \label{figure10S}
  \end{figure*}

  In order to show furthermore the agreement between the theoretical expectation and the Monte Carlo simulations in Fig.~$\ref{figure10S}$ we compare the amplitude of the  period-2 oscillations of the order parameter obtained with the Monte Carlo simulations with the predicted amplitude of the period-2 oscillations of the order parameter in a region of phase space where only period two oscillations are predicted. We find excellent agreement for both Poisson structural networks and scale-free structural networks with structural degree of the nodes uncorrelated with the regulatory degrees of the nodes.  
  \section{Further information about the real datasets investigated in the main text}
\begin{figure*}[!htb]
  \includegraphics[width=\textwidth]{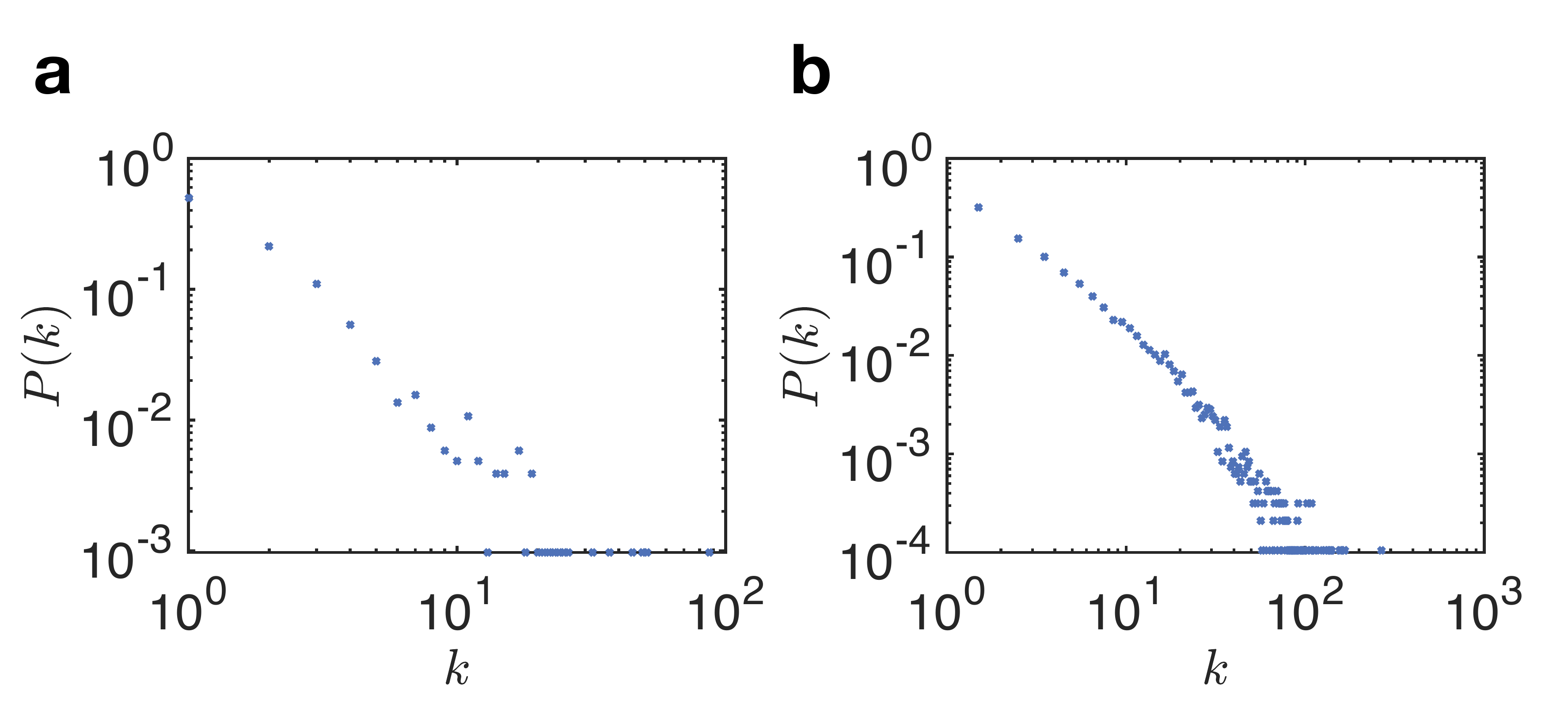}
  \caption{ {Degree distribution $P(k)$ of structural mouse brain network (a) and structural Human bio grid network (b) from \cite{Repository_networks_SI}.}}
  \label{figure11S}
 \end{figure*}

   We provide further information about the two datasets studied in Fig. $\ref{figure5}$ of the main text. In Supplementary Table $\ref{table1}$ we provide the major structural properties of the networks and in Fig.~$\ref{figure11S}$ we report their degree distribution. 

 \begin{table*}
\setlength{\tabcolsep}{5mm}{
\begin{tabular}{c c c c c c c c } 
\hline
\hline
Network & $C$ & ~$k_{min}$~ & ~$k_{max}$~ & ~$N$~ ~& ~$L$~& ~$R$~\\ 
\hline
Mouse Brain \cite{Repository_networks_SI} & ~0~ ~ & ~1~ & ~123~ & ~1029~ & ~1559~ & ~0.9592~ \\ 	
Human bio  grid \cite{Repository_networks_SI} & ~0.1612~ &  ~1~ & ~308~ & ~9436~ & ~31182~ & ~0.9642~ \\
\hline
\end{tabular}}
\caption{ {Structural properties of the real-world networks:  the averaged clustering coefficient $C$, minimum degree $k_{min}$, maximum degree $k_{max}$, number of nodes $N$, number of links $L$ and the fraction of node in the giant component $R$. Both networks are treated as undirected networks.}}
\label{table1}
\end{table*}
{ 
\section{Triadic percolation  with time delays}
In this section we provide the  equations determining triadic percolation with time delays. In presence of time delays each regulatory link  is assigned a time delay $\tau$ and Step 2 of the triadic percolation is modified by taking into account these time delays (see discussion in the main body of the paper).
We have considered two variants of triadic percolation with time delays which depend on the choice of the probability distribution for time delays of regulatory links:
\begin{itemize}
\item{[Model 1]} In Model 1 each structural link is regulated by links associated to the same time delay $\tau$ with the time delay $\tau$ being drawn from the distribution $\tilde{p}(\tau)$.
\item{[Model 2]} In Model 2 each regulatory link is associated to a time delay drawn independently from the distribution $\tilde{p}(\tau)$.
\end{itemize} 
Both models lead to a route to chaos although this dynamics is in general not in the universality class of the logistic map.

Let us discuss  how the equations determining the triadic percolation map are modified for Model 1 and Model 2 of triadic percolation with delay.
We will focus on the case of uncorrelated structural and regulatory degree of the nodes with Poisson distributions $P(\hat{\kappa}^{\pm})$.

In this case the equations for triadic percolation with delay remain Eq. (\ref{reg_percolationS2_a}), Eq. (\ref{reg_percolationS2_b}) but Eq. (\ref{reg_percolationS2_c}) is modified to take into account the delay of the regulatory interactions.
In particular in Model 1 Eq. (\ref{reg_percolationS2_c}) is substituted by  
\bea
p_L^{(t)} = p \sum_{\tau=1}^d \tilde{p}(\tau) e^{-c^{-} R^{(t+1-
\tau)}} \left[1-e^{-c^{+} R^{(t+1-\tau)}}\right]
\label{eq:delay1}
\eea
which takes into account that every structural link has all its regulatory links associated with the same delay $\tau$ where $\tau$ is draw from the distribution $\tilde{p}(\tau)$ for any structural link.

In the case in which Model 2 is considered,  each regulatory interaction  is associated to a delay $\tau$ with probability $\tilde{p}(\tau)$. Therefore among the  $\hat{\kappa}^{\pm}=\hat{\kappa}$ positive or negative regulators of a link, the probability that $n_i$ regulatory links are associated to a  delay $\tau_i$ follows a multinomial distribution
\bea
\Pi(\{n_i\}_{i=1,2,\cdots,d}|\hat{\kappa}, \tilde{\bf p}) = \frac{\hat{\kappa}!}{\prod_{i=1}^d n_i!} \prod_{i=1}^d [\tilde{p}(\tau_i)]^{n_i},
\eea
with $\tilde{\bf p}=(\tilde{p}(\tau_1),\tilde{p}(\tau_2),\ldots, \tilde{p}(\tau_d))$ and such that $\sum_{i=1}^d\tilde{p}(\tau_i)=1$.
Thus, Eq. (\ref{reg_percolationS2_c}) for  triadic percolation is modified to 
\bea
p_L^{(t)} = p \exp\left(-c^{-}\sum_{\tau_i}\tilde{p}(\tau_i)R^{(t+1-\tau_i)}\right) \left[1-\exp\left(-c^{+}\sum_{\tau_i}\tilde{p}(\tau_i)R^{(t+1-\tau_i)}\right)\right].
\label{eq:delay2}
\eea

%\subsection{Derivations of the equations determining the dynamics of model 2}

For both cases above, if we consider one step delay, \ie
\bea
\tilde{p}(\tau) = \delta_{\tau, 1},
\eea
Eq. (\ref{eq:delay1}) and Eq. (\ref{eq:delay2}) they both reduce to Eq. (\ref{reg_percolationS2_c}).

\section{Regulation of the  nodes's activity}
In this work we have proposed the model of triadic percolation that demonstrates  the important role of triadic interactions occurring in many real-world domains and whose role is related to the emergence of a time-dependent giant component.
This model is motivated by the wide spread occurrence of triadic interactions in biological, chemical and climate applications.

However and interesting variation of the triadic percolation is to consider regulatory interactions that are regulating the activity of the nodes.
In this case Step 2 of the algorithm can be modified as in the following.
\begin{itemize} 
\item[Step 2$^{\prime\prime}$] 
Each node is deactivated if at least one of the following conditions is met:
\begin{itemize}
\item[(a)] each of its positive regulatory links is connected to a  node that is inactive at Step 1;
\item[(b)] at least one of its negative regulator links  is connected to a node that is active at Step 1.
\item[(c)] neither condition (a) or (b) is met but stochastic deactivation occurs with probability $q=1-p$.
\end{itemize}

\end{itemize}

In this case, assuming uncorrelated structural and regulatory degrees the dynamics will be determined by the following equations
\bea
S^{(t)}&=&p_L^{(t-1)}\left(1-G_1\left(1-S^{(t)}\right)\right), \nonumber\\
R^{(t)}&=&p_L^{(t-1)}\left(1-G_0\left(1-S^{(t)}\right)\right), \nonumber\\
p_L^{(t)} &=& p G_0^{-}\left(1-R^{(t)}\right)\left[1-G_0^{+}\left(1-R^{(t)}\right)\right].
\eea
Interesting this model  displays period doubling and a  route to chaos of the order parameter as well (see Figure $\ref{fig:node}$). It can be directly shown, using an argument similar to the one used for triadic percolation, that the route to chaos observed in this model is in the same universality class of the logistic map as soon as $P(\hat{\kappa}^{\pm})$ indicating the distribution of the in-regulatory degree of the nodes with sign $\pm$ are Poisson distributed.
\begin{figure}[t]
  \includegraphics[width=\linewidth]{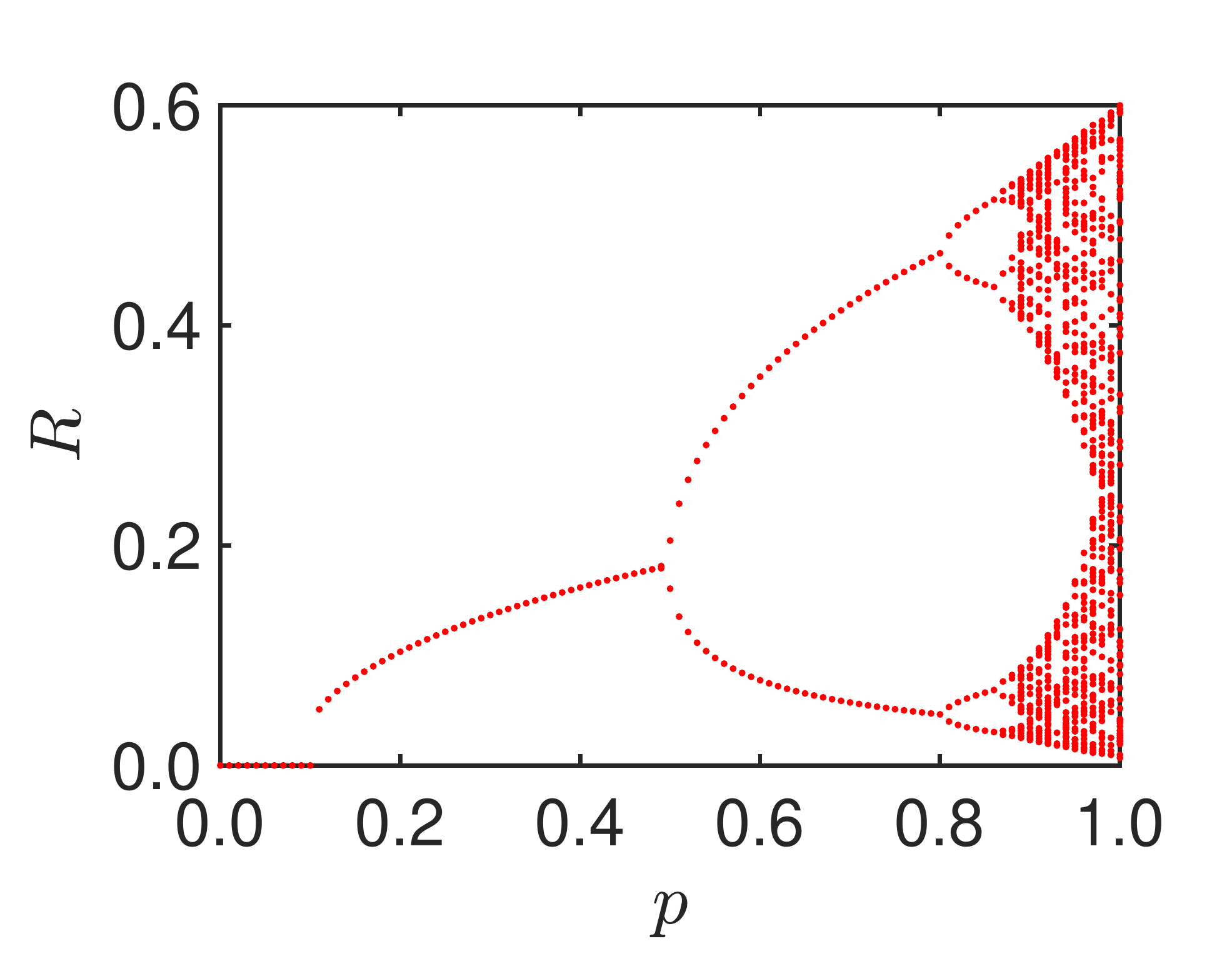}
  \caption{{ Phase diagram of percolation with regulation of the nodes's activity. The structural network is a Poisson network with an average degree $c=30$ and the regulatory in-degree distributions $P(\hat{\kappa}^{\pm}$ is also Poisson distributed with average $c^{+}=30$ and $c^{-}=5.5$. The phase diagram also displays period doubling and a route to chaos in the universality class of the route the logistic map.}}
  \label{fig:node}
\end{figure}

\section*{Supplementary References}


\begin{thebibliography}{}
\expandafter\ifx\csname url\endcsname\relax
  \def\url#1{\texttt{#1}}\fi
\expandafter\ifx\csname urlprefix\endcsname\relax\def\urlprefix{URL }\fi
\providecommand{\bibinfo}[2]{#2}
\providecommand{\eprint}[2][]{\url{#2}}

\end{thebibliography}


\begin{thebibliography}{10}
\expandafter\ifx\csname url\endcsname\relax
  \def\url#1{\texttt{#1}}\fi
\expandafter\ifx\csname urlprefix\endcsname\relax\def\urlprefix{URL }\fi
\providecommand{\bibinfo}[2]{#2}
\providecommand{\eprint}[2][]{\url{#2}}

\bibitem{dorogovtsev2008critical}
\bibinfo{author}{Dorogovtsev, S.~N.}, \bibinfo{author}{Goltsev, A.~V.} \&
  \bibinfo{author}{Mendes, J.~F.}
\newblock \bibinfo{title}{Critical phenomena in complex networks}.
\newblock \emph{\bibinfo{journal}{Reviews of Modern Physics}}
  \textbf{\bibinfo{volume}{80}}, \bibinfo{pages}{1275} (\bibinfo{year}{2008}).

\bibitem{stauffer1992introduction}
\bibinfo{author}{Stauffer, D.} \& \bibinfo{author}{Aharony, A.}
\newblock \emph{\bibinfo{title}{Introduction to percolation theory}}
  (\bibinfo{publisher}{CRC press}, \bibinfo{year}{1992}).

\bibitem{li2021percolation}
\bibinfo{author}{Li, M.} \emph{et~al.}
\newblock \bibinfo{title}{Percolation on complex networks: Theory and
  application}.
\newblock \emph{\bibinfo{journal}{Physics Reports}}
  \textbf{\bibinfo{volume}{907}}, \bibinfo{pages}{1--68}
  (\bibinfo{year}{2021}).

\bibitem{araujo2014recent}
\bibinfo{author}{Ara{\'u}jo, N.}, \bibinfo{author}{Grassberger, P.},
  \bibinfo{author}{Kahng, B.}, \bibinfo{author}{Schrenk, K.} \&
  \bibinfo{author}{Ziff, R.~M.}
\newblock \bibinfo{title}{Recent advances and open challenges in percolation}.
\newblock \emph{\bibinfo{journal}{The European Physical Journal Special
  Topics}} \textbf{\bibinfo{volume}{223}}, \bibinfo{pages}{2307--2321}
  (\bibinfo{year}{2014}).

\bibitem{buldyrev2010catastrophic}
\bibinfo{author}{Buldyrev, S.~V.}, \bibinfo{author}{Parshani, R.},
  \bibinfo{author}{Paul, G.}, \bibinfo{author}{Stanley, H.~E.} \&
  \bibinfo{author}{Havlin, S.}
\newblock \bibinfo{title}{Catastrophic cascade of failures in interdependent
  networks}.
\newblock \emph{\bibinfo{journal}{Nature}} \textbf{\bibinfo{volume}{464}},
  \bibinfo{pages}{1025--1028} (\bibinfo{year}{2010}).

\bibitem{goh2014network}
\bibinfo{author}{Min, B.}, \bibinfo{author}{Do~Yi, S.}, \bibinfo{author}{Lee,
  K.-M.} \& \bibinfo{author}{Goh, K.-I.}
\newblock \bibinfo{title}{Network robustness of multiplex networks with
  interlayer degree correlations}.
\newblock \emph{\bibinfo{journal}{Physical Review E}}
  \textbf{\bibinfo{volume}{89}}, \bibinfo{pages}{042811}
  (\bibinfo{year}{2014}).

\bibitem{kahng009percolation}
\bibinfo{author}{Cho, Y.~S.}, \bibinfo{author}{Kim, J.~S.},
  \bibinfo{author}{Park, J.}, \bibinfo{author}{Kahng, B.} \&
  \bibinfo{author}{Kim, D.}
\newblock \bibinfo{title}{Percolation transitions in scale-free networks under
  the achlioptas process}.
\newblock \emph{\bibinfo{journal}{Physical Review Letters}}
  \textbf{\bibinfo{volume}{103}}, \bibinfo{pages}{135702}
  (\bibinfo{year}{2009}).

\bibitem{boettcher2012ordinary}
\bibinfo{author}{Boettcher, S.}, \bibinfo{author}{Singh, V.} \&
  \bibinfo{author}{Ziff, R.~M.}
\newblock \bibinfo{title}{Ordinary percolation with discontinuous transitions}.
\newblock \emph{\bibinfo{journal}{Nature communications}}
  \textbf{\bibinfo{volume}{3}}, \bibinfo{pages}{1--5} (\bibinfo{year}{2012}).

\bibitem{d2019explosive}
\bibinfo{author}{D'Souza, R.~M.}, \bibinfo{author}{G{\'o}mez-Gardenes, J.},
  \bibinfo{author}{Nagler, J.} \& \bibinfo{author}{Arenas, A.}
\newblock \bibinfo{title}{Explosive phenomena in complex networks}.
\newblock \emph{\bibinfo{journal}{Advances in Physics}}
  \textbf{\bibinfo{volume}{68}}, \bibinfo{pages}{123--223}
  (\bibinfo{year}{2019}).

\bibitem{achlioptas2009explosive}
\bibinfo{author}{Achlioptas, D.}, \bibinfo{author}{D'Souza, R.~M.} \&
  \bibinfo{author}{Spencer, J.}
\newblock \bibinfo{title}{Explosive percolation in random networks}.
\newblock \emph{\bibinfo{journal}{science}} \textbf{\bibinfo{volume}{323}},
  \bibinfo{pages}{1453--1455} (\bibinfo{year}{2009}).

\bibitem{riordan2011explosive}
\bibinfo{author}{Riordan, O.} \& \bibinfo{author}{Warnke, L.}
\newblock \bibinfo{title}{Explosive percolation is continuous}.
\newblock \emph{\bibinfo{journal}{Science}} \textbf{\bibinfo{volume}{333}},
  \bibinfo{pages}{322--324} (\bibinfo{year}{2011}).

\bibitem{da2010explosive}
\bibinfo{author}{da~Costa, R.~A.}, \bibinfo{author}{Dorogovtsev, S.~N.},
  \bibinfo{author}{Goltsev, A.~V.} \& \bibinfo{author}{Mendes, J. F.~F.}
\newblock \bibinfo{title}{Explosive percolation transition is actually
  continuous}.
\newblock \emph{\bibinfo{journal}{Physical Review Letters}}
  \textbf{\bibinfo{volume}{105}}, \bibinfo{pages}{255701}
  (\bibinfo{year}{2010}).

\bibitem{nagler2020universal}
\bibinfo{author}{Fan, J.} \emph{et~al.}
\newblock \bibinfo{title}{Universal gap scaling in percolation}.
\newblock \emph{\bibinfo{journal}{Nature Physics}}
  \textbf{\bibinfo{volume}{16}}, \bibinfo{pages}{455--461}
  (\bibinfo{year}{2020}).

\bibitem{cho2013avoiding}
\bibinfo{author}{Cho, Y.~S.}, \bibinfo{author}{Hwang, S.},
  \bibinfo{author}{Herrmann, H.~J.} \& \bibinfo{author}{Kahng, B.}
\newblock \bibinfo{title}{Avoiding a spanning cluster in percolation models}.
\newblock \emph{\bibinfo{journal}{Science}} \textbf{\bibinfo{volume}{339}},
  \bibinfo{pages}{1185--1187} (\bibinfo{year}{2013}).

\bibitem{baxter2012avalanche}
\bibinfo{author}{Baxter, G.}, \bibinfo{author}{Dorogovtsev, S.},
  \bibinfo{author}{Goltsev, A.} \& \bibinfo{author}{Mendes, J.}
\newblock \bibinfo{title}{Avalanche collapse of interdependent networks}.
\newblock \emph{\bibinfo{journal}{Physical Review Letters}}
  \textbf{\bibinfo{volume}{109}}, \bibinfo{pages}{248701}
  (\bibinfo{year}{2012}).

\bibitem{radicchi2015percolation}
\bibinfo{author}{Radicchi, F.}
\newblock \bibinfo{title}{Percolation in real interdependent networks}.
\newblock \emph{\bibinfo{journal}{Nature Physics}}
  \textbf{\bibinfo{volume}{11}}, \bibinfo{pages}{597--602}
  (\bibinfo{year}{2015}).

\bibitem{radicchi2017redundant}
\bibinfo{author}{Radicchi, F.} \& \bibinfo{author}{Bianconi, G.}
\newblock \bibinfo{title}{Redundant interdependencies boost the robustness of
  multiplex networks}.
\newblock \emph{\bibinfo{journal}{Physical Review X}}
  \textbf{\bibinfo{volume}{7}}, \bibinfo{pages}{011013} (\bibinfo{year}{2017}).

\bibitem{reis2014avoiding}
\bibinfo{author}{Reis, S.~D.} \emph{et~al.}
\newblock \bibinfo{title}{Avoiding catastrophic failure in correlated networks
  of networks}.
\newblock \emph{\bibinfo{journal}{Nature Physics}}
  \textbf{\bibinfo{volume}{10}}, \bibinfo{pages}{762--767}
  (\bibinfo{year}{2014}).

\bibitem{kryven2019bond}
\bibinfo{author}{Kryven, I.}
\newblock \bibinfo{title}{Bond percolation in coloured and multiplex networks}.
\newblock \emph{\bibinfo{journal}{Nature communications}}
  \textbf{\bibinfo{volume}{10}}, \bibinfo{pages}{1--16} (\bibinfo{year}{2019}).

\bibitem{gao2012networks}
\bibinfo{author}{Gao, J.}, \bibinfo{author}{Buldyrev, S.~V.},
  \bibinfo{author}{Stanley, H.~E.} \& \bibinfo{author}{Havlin, S.}
\newblock \bibinfo{title}{Networks formed from interdependent networks}.
\newblock \emph{\bibinfo{journal}{Nature physics}}
  \textbf{\bibinfo{volume}{8}}, \bibinfo{pages}{40--48} (\bibinfo{year}{2012}).

\bibitem{bianconi2018multilayer}
\bibinfo{author}{Bianconi, G.}
\newblock \emph{\bibinfo{title}{Multilayer networks: structure and function}}
  (\bibinfo{publisher}{Oxford university press}, \bibinfo{year}{2018}).

\bibitem{boccaletti2014structure}
\bibinfo{author}{Boccaletti, S.} \emph{et~al.}
\newblock \bibinfo{title}{The structure and dynamics of multilayer networks}.
\newblock \emph{\bibinfo{journal}{Physics reports}}
  \textbf{\bibinfo{volume}{544}}, \bibinfo{pages}{1--122}
  (\bibinfo{year}{2014}).

\bibitem{kivela2014multilayer}
\bibinfo{author}{Kivel{\"a}, M.} \emph{et~al.}
\newblock \bibinfo{title}{Multilayer networks}.
\newblock \emph{\bibinfo{journal}{Journal of complex networks}}
  \textbf{\bibinfo{volume}{2}}, \bibinfo{pages}{203--271}
  (\bibinfo{year}{2014}).

\bibitem{zhao2013antagonistic}
\bibinfo{author}{Zhao, K.} \& \bibinfo{author}{Bianconi, G.}
\newblock \bibinfo{title}{Percolation on interacting, antagonistic networks}.
\newblock \emph{\bibinfo{journal}{Journal of Statistical Mechanics: Theory and
  Experiment}} \textbf{\bibinfo{volume}{2013}}, \bibinfo{pages}{P05005}
  (\bibinfo{year}{2013}).

\bibitem{danziger2019dynamic}
\bibinfo{author}{Danziger, M.~M.}, \bibinfo{author}{Bonamassa, I.},
  \bibinfo{author}{Boccaletti, S.} \& \bibinfo{author}{Havlin, S.}
\newblock \bibinfo{title}{Dynamic interdependence and competition in multilayer
  networks}.
\newblock \emph{\bibinfo{journal}{Nature Physics}}
  \textbf{\bibinfo{volume}{15}}, \bibinfo{pages}{178--185}
  (\bibinfo{year}{2019}).

\bibitem{shekhtman2016recent}
\bibinfo{author}{Shekhtman, L.~M.}, \bibinfo{author}{Danziger, M.~M.} \&
  \bibinfo{author}{Havlin, S.}
\newblock \bibinfo{title}{Recent advances on failure and recovery in networks
  of networks}.
\newblock \emph{\bibinfo{journal}{Chaos, Solitons \& Fractals}}
  \textbf{\bibinfo{volume}{90}}, \bibinfo{pages}{28--36}
  (\bibinfo{year}{2016}).

\bibitem{watanabe2016resilience}
\bibinfo{author}{Watanabe, S.} \& \bibinfo{author}{Kabashima, Y.}
\newblock \bibinfo{title}{Resilience of antagonistic networks with regard to
  the effects of initial failures and degree-degree correlations}.
\newblock \emph{\bibinfo{journal}{Physical Review E}}
  \textbf{\bibinfo{volume}{94}}, \bibinfo{pages}{032308}
  (\bibinfo{year}{2016}).

\bibitem{kotnis2015percolation}
\bibinfo{author}{Kotnis, B.} \& \bibinfo{author}{Kuri, J.}
\newblock \bibinfo{title}{Percolation on networks with antagonistic and
  dependent interactions}.
\newblock \emph{\bibinfo{journal}{Physical Review E}}
  \textbf{\bibinfo{volume}{91}}, \bibinfo{pages}{032805}
  (\bibinfo{year}{2015}).

\bibitem{majdandzic2016multiple}
\bibinfo{author}{Majdandzic, A.} \emph{et~al.}
\newblock \bibinfo{title}{Multiple tipping points and optimal repairing in
  interacting networks}.
\newblock \emph{\bibinfo{journal}{Nature communications}}
  \textbf{\bibinfo{volume}{7}}, \bibinfo{pages}{1--10} (\bibinfo{year}{2016}).

\bibitem{danziger2022recovery}
\bibinfo{author}{Danziger, M.~M.} \& \bibinfo{author}{Barab{\'a}si, A.-L.}
\newblock \bibinfo{title}{Recovery coupling in multilayer networks}.
\newblock \emph{\bibinfo{journal}{Nature communications}}
  \textbf{\bibinfo{volume}{13}}, \bibinfo{pages}{1--8} (\bibinfo{year}{2022}).

\bibitem{battiston2020networks}
\bibinfo{author}{Battiston, F.} \emph{et~al.}
\newblock \bibinfo{title}{Networks beyond pairwise interactions: structure and
  dynamics}.
\newblock \emph{\bibinfo{journal}{Physics Reports}}
  \textbf{\bibinfo{volume}{874}}, \bibinfo{pages}{1--92}
  (\bibinfo{year}{2020}).

\bibitem{bianconi2021higher}
\bibinfo{author}{Bianconi, G.}
\newblock \emph{\bibinfo{title}{Higher-order networks: An Introduction to
  Simplicial Complexes}} (\bibinfo{publisher}{Cambridge University Press},
  \bibinfo{year}{2021}).

\bibitem{perc2022dynamics}
\bibinfo{author}{Majhi, S.}, \bibinfo{author}{Perc, M.} \&
  \bibinfo{author}{Ghosh, D.}
\newblock \bibinfo{title}{Dynamics on higher-order networks: A review}.
\newblock \emph{\bibinfo{journal}{Journal of the Royal Society Interface}}
  \textbf{\bibinfo{volume}{19}}, \bibinfo{pages}{20220043}
  (\bibinfo{year}{2022}).

\bibitem{lambiotte2018simplicial}
\bibinfo{author}{Salnikov, V.}, \bibinfo{author}{Cassese, D.} \&
  \bibinfo{author}{Lambiotte, R.}
\newblock \bibinfo{title}{Simplicial complexes and complex systems}.
\newblock \emph{\bibinfo{journal}{European Journal of Physics}}
  \textbf{\bibinfo{volume}{40}}, \bibinfo{pages}{014001}
  (\bibinfo{year}{2018}).

\bibitem{bick2021higher}
\bibinfo{author}{Bick, C.}, \bibinfo{author}{Gross, E.},
  \bibinfo{author}{Harrington, H.~A.} \& \bibinfo{author}{Schaub, M.~T.}
\newblock \bibinfo{title}{What are higher-order networks?}
\newblock \emph{\bibinfo{journal}{arXiv preprint arXiv:2104.11329}}
  (\bibinfo{year}{2021}).

\bibitem{torres2021and}
\bibinfo{author}{Torres, L.}, \bibinfo{author}{Blevins, A.~S.},
  \bibinfo{author}{Bassett, D.} \& \bibinfo{author}{Eliassi-Rad, T.}
\newblock \bibinfo{title}{The why, how, and when of representations for complex
  systems}.
\newblock \emph{\bibinfo{journal}{SIAM Review}} \textbf{\bibinfo{volume}{63}},
  \bibinfo{pages}{435--485} (\bibinfo{year}{2021}).

\bibitem{giusti2016two}
\bibinfo{author}{Giusti, C.}, \bibinfo{author}{Ghrist, R.} \&
  \bibinfo{author}{Bassett, D.~S.}
\newblock \bibinfo{title}{Two’s company, three (or more) is a simplex}.
\newblock \emph{\bibinfo{journal}{Journal of computational neuroscience}}
  \textbf{\bibinfo{volume}{41}}, \bibinfo{pages}{1--14} (\bibinfo{year}{2016}).

\bibitem{faskowitz2022edges}
\bibinfo{author}{Faskowitz, J.}, \bibinfo{author}{Betzel, R.~F.} \&
  \bibinfo{author}{Sporns, O.}
\newblock \bibinfo{title}{Edges in brain networks: Contributions to models of
  structure and function}.
\newblock \emph{\bibinfo{journal}{Network Neuroscience}}
  \textbf{\bibinfo{volume}{6}}, \bibinfo{pages}{1--28} (\bibinfo{year}{2022}).

\bibitem{jost2019hypergraph}
\bibinfo{author}{Jost, J.} \& \bibinfo{author}{Mulas, R.}
\newblock \bibinfo{title}{Hypergraph laplace operators for chemical reaction
  networks}.
\newblock \emph{\bibinfo{journal}{Advances in mathematics}}
  \textbf{\bibinfo{volume}{351}}, \bibinfo{pages}{870--896}
  (\bibinfo{year}{2019}).

\bibitem{boers2019complex}
\bibinfo{author}{Boers, N.} \emph{et~al.}
\newblock \bibinfo{title}{Complex networks reveal global pattern of
  extreme-rainfall teleconnections}.
\newblock \emph{\bibinfo{journal}{Nature}} \textbf{\bibinfo{volume}{566}},
  \bibinfo{pages}{373--377} (\bibinfo{year}{2019}).

\bibitem{su2022climatic}
\bibinfo{author}{Su, Z.}, \bibinfo{author}{Meyerhenke, H.} \&
  \bibinfo{author}{Kurths, J.}
\newblock \bibinfo{title}{The climatic interdependence of extreme-rainfall
  events around the globe}.
\newblock \emph{\bibinfo{journal}{Chaos: An Interdisciplinary Journal of
  Nonlinear Science}} \textbf{\bibinfo{volume}{32}}, \bibinfo{pages}{043126}
  (\bibinfo{year}{2022}).

\bibitem{millan2020explosive}
\bibinfo{author}{Mill{\'a}n, A.~P.}, \bibinfo{author}{Torres, J.~J.} \&
  \bibinfo{author}{Bianconi, G.}
\newblock \bibinfo{title}{Explosive higher-order kuramoto dynamics on
  simplicial complexes}.
\newblock \emph{\bibinfo{journal}{Physical Review Letters}}
  \textbf{\bibinfo{volume}{124}}, \bibinfo{pages}{218301}
  (\bibinfo{year}{2020}).

\bibitem{skardal2019abrupt}
\bibinfo{author}{Skardal, P.~S.} \& \bibinfo{author}{Arenas, A.}
\newblock \bibinfo{title}{Abrupt desynchronization and extensive multistability
  in globally coupled oscillator simplexes}.
\newblock \emph{\bibinfo{journal}{Physical Review Letters}}
  \textbf{\bibinfo{volume}{122}}, \bibinfo{pages}{248301}
  (\bibinfo{year}{2019}).

\bibitem{zhang2021unified}
\bibinfo{author}{Zhang, Y.}, \bibinfo{author}{Latora, V.} \&
  \bibinfo{author}{Motter, A.~E.}
\newblock \bibinfo{title}{Unified treatment of synchronization patterns in
  generalized networks with higher-order, multilayer, and temporal
  interactions}.
\newblock \emph{\bibinfo{journal}{Communications Physics}}
  \textbf{\bibinfo{volume}{4}}, \bibinfo{pages}{1--9} (\bibinfo{year}{2021}).

\bibitem{mulas2020coupled}
\bibinfo{author}{Mulas, R.}, \bibinfo{author}{Kuehn, C.} \&
  \bibinfo{author}{Jost, J.}
\newblock \bibinfo{title}{Coupled dynamics on hypergraphs: Master stability of
  steady states and synchronization}.
\newblock \emph{\bibinfo{journal}{Physical Review E}}
  \textbf{\bibinfo{volume}{101}}, \bibinfo{pages}{062313}
  (\bibinfo{year}{2020}).

\bibitem{carletti2020random}
\bibinfo{author}{Carletti, T.}, \bibinfo{author}{Battiston, F.},
  \bibinfo{author}{Cencetti, G.} \& \bibinfo{author}{Fanelli, D.}
\newblock \bibinfo{title}{Random walks on hypergraphs}.
\newblock \emph{\bibinfo{journal}{Physical review E}}
  \textbf{\bibinfo{volume}{101}}, \bibinfo{pages}{022308}
  (\bibinfo{year}{2020}).

\bibitem{st2021universal}
\bibinfo{author}{St-Onge, G.}, \bibinfo{author}{Sun, H.},
  \bibinfo{author}{Allard, A.}, \bibinfo{author}{H{\'e}bert-Dufresne, L.} \&
  \bibinfo{author}{Bianconi, G.}
\newblock \bibinfo{title}{Universal nonlinear infection kernel from
  heterogeneous exposure on higher-order networks}.
\newblock \emph{\bibinfo{journal}{Physical Review Letters}}
  \textbf{\bibinfo{volume}{127}}, \bibinfo{pages}{158301}
  (\bibinfo{year}{2021}).

\bibitem{de2020social}
\bibinfo{author}{de~Arruda, G.~F.}, \bibinfo{author}{Petri, G.} \&
  \bibinfo{author}{Moreno, Y.}
\newblock \bibinfo{title}{Social contagion models on hypergraphs}.
\newblock \emph{\bibinfo{journal}{Physical Review Research}}
  \textbf{\bibinfo{volume}{2}}, \bibinfo{pages}{023032} (\bibinfo{year}{2020}).

\bibitem{iacopini2019simplicial}
\bibinfo{author}{Iacopini, I.}, \bibinfo{author}{Petri, G.},
  \bibinfo{author}{Barrat, A.} \& \bibinfo{author}{Latora, V.}
\newblock \bibinfo{title}{Simplicial models of social contagion}.
\newblock \emph{\bibinfo{journal}{Nature communications}}
  \textbf{\bibinfo{volume}{10}}, \bibinfo{pages}{1--9} (\bibinfo{year}{2019}).

\bibitem{ferraz2021phase}
\bibinfo{author}{Ferraz~de Arruda, G.}, \bibinfo{author}{Tizzani, M.} \&
  \bibinfo{author}{Moreno, Y.}
\newblock \bibinfo{title}{Phase transitions and stability of dynamical
  processes on hypergraphs}.
\newblock \emph{\bibinfo{journal}{Communications Physics}}
  \textbf{\bibinfo{volume}{4}}, \bibinfo{pages}{1--9} (\bibinfo{year}{2021}).

\bibitem{sun2021higher}
\bibinfo{author}{Sun, H.} \& \bibinfo{author}{Bianconi, G.}
\newblock \bibinfo{title}{Higher-order percolation processes on multiplex
  hypergraphs}.
\newblock \emph{\bibinfo{journal}{Physical Review E}}
  \textbf{\bibinfo{volume}{104}}, \bibinfo{pages}{034306}
  (\bibinfo{year}{2021}).

\bibitem{taylor2015topological}
\bibinfo{author}{Taylor, D.} \emph{et~al.}
\newblock \bibinfo{title}{Topological data analysis of contagion maps for
  examining spreading processes on networks}.
\newblock \emph{\bibinfo{journal}{Nature communications}}
  \textbf{\bibinfo{volume}{6}}, \bibinfo{pages}{1--11} (\bibinfo{year}{2015}).

\bibitem{alvarez2021evolutionary}
\bibinfo{author}{Alvarez-Rodriguez, U.} \emph{et~al.}
\newblock \bibinfo{title}{Evolutionary dynamics of higher-order interactions in
  social networks}.
\newblock \emph{\bibinfo{journal}{Nature Human Behaviour}}
  \textbf{\bibinfo{volume}{5}}, \bibinfo{pages}{586--595}
  (\bibinfo{year}{2021}).

\bibitem{bianconi2018topological}
\bibinfo{author}{Bianconi, G.} \& \bibinfo{author}{Ziff, R.~M.}
\newblock \bibinfo{title}{Topological percolation on hyperbolic simplicial
  complexes}.
\newblock \emph{\bibinfo{journal}{Physical Review E}}
  \textbf{\bibinfo{volume}{98}}, \bibinfo{pages}{052308}
  (\bibinfo{year}{2018}).

\bibitem{lee2021homological}
\bibinfo{author}{Lee, Y.}, \bibinfo{author}{Lee, J.}, \bibinfo{author}{Oh,
  S.~M.}, \bibinfo{author}{Lee, D.} \& \bibinfo{author}{Kahng, B.}
\newblock \bibinfo{title}{Homological percolation transitions in growing
  simplicial complexes}.
\newblock \emph{\bibinfo{journal}{Chaos: An Interdisciplinary Journal of
  Nonlinear Science}} \textbf{\bibinfo{volume}{31}}, \bibinfo{pages}{041102}
  (\bibinfo{year}{2021}).

\bibitem{bianconi2019percolation}
\bibinfo{author}{Bianconi, G.}, \bibinfo{author}{Kryven, I.} \&
  \bibinfo{author}{Ziff, R.~M.}
\newblock \bibinfo{title}{Percolation on branching simplicial and cell
  complexes and its relation to interdependent percolation}.
\newblock \emph{\bibinfo{journal}{Physical Review E}}
  \textbf{\bibinfo{volume}{100}}, \bibinfo{pages}{062311}
  (\bibinfo{year}{2019}).

\bibitem{bobrowski2020homological}
\bibinfo{author}{Bobrowski, O.} \& \bibinfo{author}{Skraba, P.}
\newblock \bibinfo{title}{Homological percolation and the euler
  characteristic}.
\newblock \emph{\bibinfo{journal}{Physical Review E}}
  \textbf{\bibinfo{volume}{101}}, \bibinfo{pages}{032304}
  (\bibinfo{year}{2020}).

\bibitem{bao2022impact}
\bibinfo{author}{Bao, X.} \emph{et~al.}
\newblock \bibinfo{title}{Impact of basic network motifs on the collective
  response to perturbations}.
\newblock \emph{\bibinfo{journal}{Nature communications}}
  \textbf{\bibinfo{volume}{13}}, \bibinfo{pages}{1--8} (\bibinfo{year}{2022}).

\bibitem{kishony2016high}
\bibinfo{author}{Bairey, E.}, \bibinfo{author}{Kelsic, E.~D.} \&
  \bibinfo{author}{Kishony, R.}
\newblock \bibinfo{title}{High-order species interactions shape ecosystem
  diversity}.
\newblock \emph{\bibinfo{journal}{Nature communications}}
  \textbf{\bibinfo{volume}{7}}, \bibinfo{pages}{1--7} (\bibinfo{year}{2016}).

\bibitem{grilli2017higher}
\bibinfo{author}{Grilli, J.}, \bibinfo{author}{Barab{\'a}s, G.},
  \bibinfo{author}{Michalska-Smith, M.~J.} \& \bibinfo{author}{Allesina, S.}
\newblock \bibinfo{title}{Higher-order interactions stabilize dynamics in
  competitive network models}.
\newblock \emph{\bibinfo{journal}{Nature}} \textbf{\bibinfo{volume}{548}},
  \bibinfo{pages}{210--213} (\bibinfo{year}{2017}).

\bibitem{stouffer2019mechanistic}
\bibinfo{author}{Letten, A.~D.} \& \bibinfo{author}{Stouffer, D.~B.}
\newblock \bibinfo{title}{The mechanistic basis for higher-order interactions
  and non-additivity in competitive communities}.
\newblock \emph{\bibinfo{journal}{Ecology letters}}
  \textbf{\bibinfo{volume}{22}}, \bibinfo{pages}{423--436}
  (\bibinfo{year}{2019}).

\bibitem{cho2016optogenetic}
\bibinfo{author}{Cho, W.-H.}, \bibinfo{author}{Barcelon, E.} \&
  \bibinfo{author}{Lee, S.~J.}
\newblock \bibinfo{title}{Optogenetic glia manipulation: possibilities and
  future prospects}.
\newblock \emph{\bibinfo{journal}{Experimental neurobiology}}
  \textbf{\bibinfo{volume}{25}}, \bibinfo{pages}{197} (\bibinfo{year}{2016}).

\bibitem{arenas2008synchronization}
\bibinfo{author}{Arenas, A.}, \bibinfo{author}{D{\'\i}az-Guilera, A.},
  \bibinfo{author}{Kurths, J.}, \bibinfo{author}{Moreno, Y.} \&
  \bibinfo{author}{Zhou, C.}
\newblock \bibinfo{title}{Synchronization in complex networks}.
\newblock \emph{\bibinfo{journal}{Physics reports}}
  \textbf{\bibinfo{volume}{469}}, \bibinfo{pages}{93--153}
  (\bibinfo{year}{2008}).

\bibitem{kurths2009complex}
\bibinfo{author}{Marwan, N.}, \bibinfo{author}{Donges, J.~F.},
  \bibinfo{author}{Zou, Y.}, \bibinfo{author}{Donner, R.~V.} \&
  \bibinfo{author}{Kurths, J.}
\newblock \bibinfo{title}{Complex network approach for recurrence analysis of
  time series}.
\newblock \emph{\bibinfo{journal}{Physics Letters A}}
  \textbf{\bibinfo{volume}{373}}, \bibinfo{pages}{4246--4254}
  (\bibinfo{year}{2009}).

\bibitem{strogatz2018nonlinear}
\bibinfo{author}{Strogatz, S.~H.}
\newblock \emph{\bibinfo{title}{Nonlinear dynamics and chaos: with applications
  to physics, biology, chemistry, and engineering}} (\bibinfo{publisher}{CRC
  press}, \bibinfo{year}{2018}).

\bibitem{porter2016dynamical}
\bibinfo{author}{Porter, M.~A.} \& \bibinfo{author}{Gleeson, J.~P.}
\newblock \bibinfo{title}{Dynamical systems on networks}.
\newblock \emph{\bibinfo{journal}{Frontiers in Applied Dynamical Systems:
  Reviews and Tutorials}} \textbf{\bibinfo{volume}{4}} (\bibinfo{year}{2016}).


\bibitem{Repository_networks}
\bibinfo{author}{Rossi, R.~A.} \& \bibinfo{author}{Ahmed, N.~K.}
\newblock \bibinfo{title}{The network data repository with interactive graph
  analytics and visualization}.
\newblock In \emph{\bibinfo{booktitle}{Proceedings of the Twenty-Ninth AAAI
  Conference on Artificial Intelligence}} (\bibinfo{year}{2015}).
\newblock \urlprefix\url{http://networkrepository.com}.

\bibitem{mezard1987spin}
\bibinfo{author}{M{\'e}zard, M.}, \bibinfo{author}{Parisi, G.} \&
  \bibinfo{author}{Virasoro, M.~A.}
\newblock \emph{\bibinfo{title}{Spin glass theory and beyond: An Introduction
  to the Replica Method and Its Applications}}, vol.~\bibinfo{volume}{9}
  (\bibinfo{publisher}{World Scientific Publishing Company},
  \bibinfo{year}{1987}).

\bibitem{motter2018antagonistic}
\bibinfo{author}{Motter, A.~E.} \& \bibinfo{author}{Timme, M.}
\newblock \bibinfo{title}{Antagonistic phenomena in network dynamics}.
\newblock \emph{\bibinfo{journal}{Annual review of condensed matter physics}}
  \textbf{\bibinfo{volume}{9}}, \bibinfo{pages}{463--484}
  (\bibinfo{year}{2018}).

\end{thebibliography}

\begin{thebibliography}{10}
\expandafter\ifx\csname url\endcsname\relax
  \def\url#1{\texttt{#1}}\fi
\expandafter\ifx\csname urlprefix\endcsname\relax\def\urlprefix{URL }\fi
\providecommand{\bibinfo}[2]{#2}
\providecommand{\eprint}[2][]{\url{#2}}

\bibitem{strogatz2018nonlinear_SI}
\bibinfo{author}{Strogatz, S.~H.}
\newblock \emph{\bibinfo{title}{Nonlinear dynamics and chaos: with applications
  to physics, biology, chemistry, and engineering}} (\bibinfo{publisher}{CRC
  press}, \bibinfo{year}{2018}).
  
  
 \bibitem{feigenbaum1978quantitative_SI}
 \bibinfo{author}{Feigenbaum, M.~J.}
  \newblock {\bibinfo{title}{Quantitative universality for a class of nonlinear transformations}} \newblock \emph{\bibinfo{journal}{Journal of statistical physics}}
  \textbf{\bibinfo{volume}{19}}, \bibinfo{pages}{25-52} (\bibinfo{year}{1978}).

  
  
\bibitem{dorogovtsev2008critical_SI}
\bibinfo{author}{Dorogovtsev, S.~N.}, \bibinfo{author}{Goltsev, A.~V.} \&
  \bibinfo{author}{Mendes, J.~F.}
\newblock \bibinfo{title}{Critical phenomena in complex networks}.
\newblock \emph{\bibinfo{journal}{Reviews of Modern Physics}}
  \textbf{\bibinfo{volume}{80}}, \bibinfo{pages}{1275} (\bibinfo{year}{2008}).

\bibitem{buldyrev2010catastrophic_SI}
\bibinfo{author}{Buldyrev, S.~V.}, \bibinfo{author}{Parshani, R.},
  \bibinfo{author}{Paul, G.}, \bibinfo{author}{Stanley, H.~E.} \&
  \bibinfo{author}{Havlin, S.}
\newblock \bibinfo{title}{Catastrophic cascade of failures in interdependent
  networks}.
\newblock \emph{\bibinfo{journal}{Nature}} \textbf{\bibinfo{volume}{464}},
  \bibinfo{pages}{1025--1028} (\bibinfo{year}{2010}).

\bibitem{Repository_networks_SI}
\bibinfo{author}{Rossi, R.~A.} \& \bibinfo{author}{Ahmed, N.~K.}
\newblock \bibinfo{title}{The network data repository with interactive graph
  analytics and visualization}.
\newblock In \emph{\bibinfo{booktitle}{Proceedings of the Twenty-Ninth AAAI
  Conference on Artificial Intelligence}} (\bibinfo{year}{2015}).
\newblock \urlprefix\url{http://networkrepository.com}.


\end{thebibliography}
\end{document}